\documentclass{./packages/aa}  
\usepackage{lscape}
\usepackage{graphicx}
\usepackage{stfloats}
\usepackage[dvips]{color}
\usepackage{./bibtex/natbib}
\bibpunct{(}{)}{;}{a}{}{,}                           
\usepackage{txfonts}
\usepackage{hyperref}
\hypersetup{colorlinks=true,citecolor=black,linkcolor=black}

\begin{document}
\bibliographystyle{./bibtex/aa.bst}
   \title{Two Type Ic supernovae in low-metallicity, dwarf galaxies: diversity of explosions.}

   \subtitle{}

   \author{D. R. Young \inst{1}\thanks{email: dyoung06@qub.ac.uk}
      \and S. J. Smartt \inst{1}
      \and S. Valenti \inst{1}
      \and A. Pastorello \inst{1}
      \and S. Benetti \inst{2}
      \and C. R. Benn \inst{3}
      \and D. Bersier \inst{4}
      \and M. T.  Botticella \inst{1}
      \and R. L. M. Corradi \inst{3}
      \and A. H. Harutyunyan \inst{2}
      \and M. Hrudkova \inst{3,5}
      \and I. Hunter \inst{1}
      \and S. Mattila \inst{6}
      \and E. J. W. de Mooij \inst{7}
      \and H. Navasardyan \inst{2}
      \and I. A. G. Snellen \inst{7}
      \and N. R. Tanvir \inst{8}
      \and L. Zampieri \inst{2}}

   \institute{Astrophysics Research Centre, School of Maths and Physics, Queen's University Belfast, Belfast BT7 1NN, Northern Ireland, UK.
   \and INAF, Osservatorio Astronomico di Padova, vicolo dell'Osservatorio 5, I-35122 Padova, Italy
   \and Isaac Newton Group, Apartado 321, E-38700 Santa Cruz de La Palma, Spain
   \and Astrophysics Research Institute, Liverpool John Moores University, Twelve Quays House, Egerton Wharf, Birkenhead, CH41 1LD, UK.
   \and Astronomical Institute of the Charles University, Faculty of Mathematics and Physics, V Hole\v{s}ovi\v{c}k\'{a}ch 2, 18000, Praha 8, Czech Republic
   \and Tuorla Observatory, Department of Physics and Astronomy, University of Turku, V\"ais\"al\"antie 20, FI-21500 Piikki\"o, Finland.
   \and Leiden Observatory, Leiden University, Postbus 9513, 2300 RA, Leiden, The Netherlands
   \and Department of Physics and Astronomy, University of Leicester, Leicester, LE1 7RH, UK.}

   \date{Received October 10, 2009}
      
   \abstract
    {}
   	{The aim of this paper is to discuss the nature of two Type Ic supernovae SN 2007bg and SN 2007bi and their host galaxies. Both supernovae were discovered in wide-field, non-targeted surveys and are found to be associated with sub-luminous blue dwarf galaxies identified in SDSS images.}
   	{We present $BVRI$ photometry and optical spectroscopy of SN 2007bg and SN 2007bi and their host galaxies. Their lightcurves and spectra are compared to those of other Type Ic SNe and analysis of these data provides estimates of the energetics, total ejected masses and synthesised mass of $^{56}$Ni.  Detection of the host galaxy emission lines allows for metallicity measurements.}
   	{Neither SNe 2007bg nor 2007bi were found in association with an observed GRB, but from estimates of the metallicities of their host-galaxies they are found to inhabit similar low-metallicity environments as GRB associated supernovae. The radio-bright SN 2007bg is hosted by an extremely sub-luminous galaxy of magnitude $M_{B}=-12.4${\tiny$\pm0.6$} mag and an estimated oxygen abundance of $12+\log({\rm O/H})=8.18${\tiny $\pm0.17$} (on the \citealt{Pettini:2004p217} scale). The early lightcurve evolution of SN 2007bg matches the fast-pace decline of SN 1994I giving it one of the fastest post-maximum decline rates of all broad-lined Type Ic supernovae known to date and, when combined with its high expansion velocities, a high kinetic energy to ejected mass ratio ($E_K/M_{ej}\sim2.7$).
   We also show that SN 2007bi is possibly the most luminous Type Ic known, reaching a peak magnitude of $M_{R} \sim -21.3$ mag and  displays a remarkably slow decline, following the radioactive decay rate of $^{56}$Co to $^{56}$Fe throughout the course of its observed lifetime. SN 2007bi also displays an extreme longevity in its spectral evolution and is still not fully nebular at approximately one year post-maximum brightness. From a simple model of the bolometric light curve of SN 2007bi we estimate a total ejected $^{56}$Ni mass of $M_{Ni}=3.5-4.5{\rm M}_\odot$, the largest $^{56}$Ni mass measured in the ejecta of a supernova to date. There are two models that could explain the high luminosity and large ejected $^{56}$Ni mass. One is a pair-instability supernova (PISN) which has been predicted to occur for massive stars at low metallicities. We measure the host galaxy metallicity of SN 2007bi to be $12+\log({\rm O/H}) = 8.15${\tiny $\pm0.15$} (on the \citealt{McGaugh:1991p7985} scale) which is somewhat high to be consistent with the PISN model. An alternative is the core-collapse of a C+O star of $20-40$M$_{\odot}$ which is the core of a star of originally $50-100$M$_{\odot}$.}
   	{}
   

\titlerunning{Two Type Ic supernovae in low-metallicity, dwarf galaxies: diversity of explosions.}
\maketitle

\section{Introduction} 

The absence of obvious Hydrogen and Helium in the spectra of supernovae Type Ic (SNe Ic) indicates that these objects originate from the gravitational core-collapse of stars that have been stripped of their outer layers during the course of their evolution. This stripping is thought to occur either via binary interaction or mass-loss resulting from the strong stellar winds suffered during the life-time of the progenitor star \cite{Smartt:2009p11794}. As a class SNe Ic are known to exhibit a broad range in both luminosity and light-curve shape \citep{Richardson:2006p72}; a heterogeneity also displayed in their spectra \citep{Matheson:2001p4883,Elmhamdi:2006p6842}. Interest in this class of SN has been rejuvenated in the past decade with confirmed association of SNe Ic with long-duration Gamma-Ray Bursts (GRBs). In every case that a SN has been identified in association with a GRB it has been classified as a broad-lined SN Ic (SN Ic-BL): GRB 980425/SN 1998bw \citep{Galama:1998p4661}, GRB 030329/SN 2003dh \citep{Stanek:2003p8187, Hjorth:2003p8194, Matheson:2003p8201}, GRB 031203/SN 2003lw \citep{Malesani:2004p8254, Mazzali:2006p8264} and GRB 060218/SN 2006aj \citep{Modjaz:2006p4901, Pian:2006p8257, Mazzali:2006p8258, Campana:2006p8259, Sollerman:2006p4932, Mirabal:2006p8260, Cobb:2006p8263}. The broad-lines exhibited by these SNe are indicative of the huge photospheric velocities achieved in their explosions.

\begin{figure*}		
\begin{minipage}[t]{0.48\linewidth}		
\centering
\includegraphics[scale=0.52]{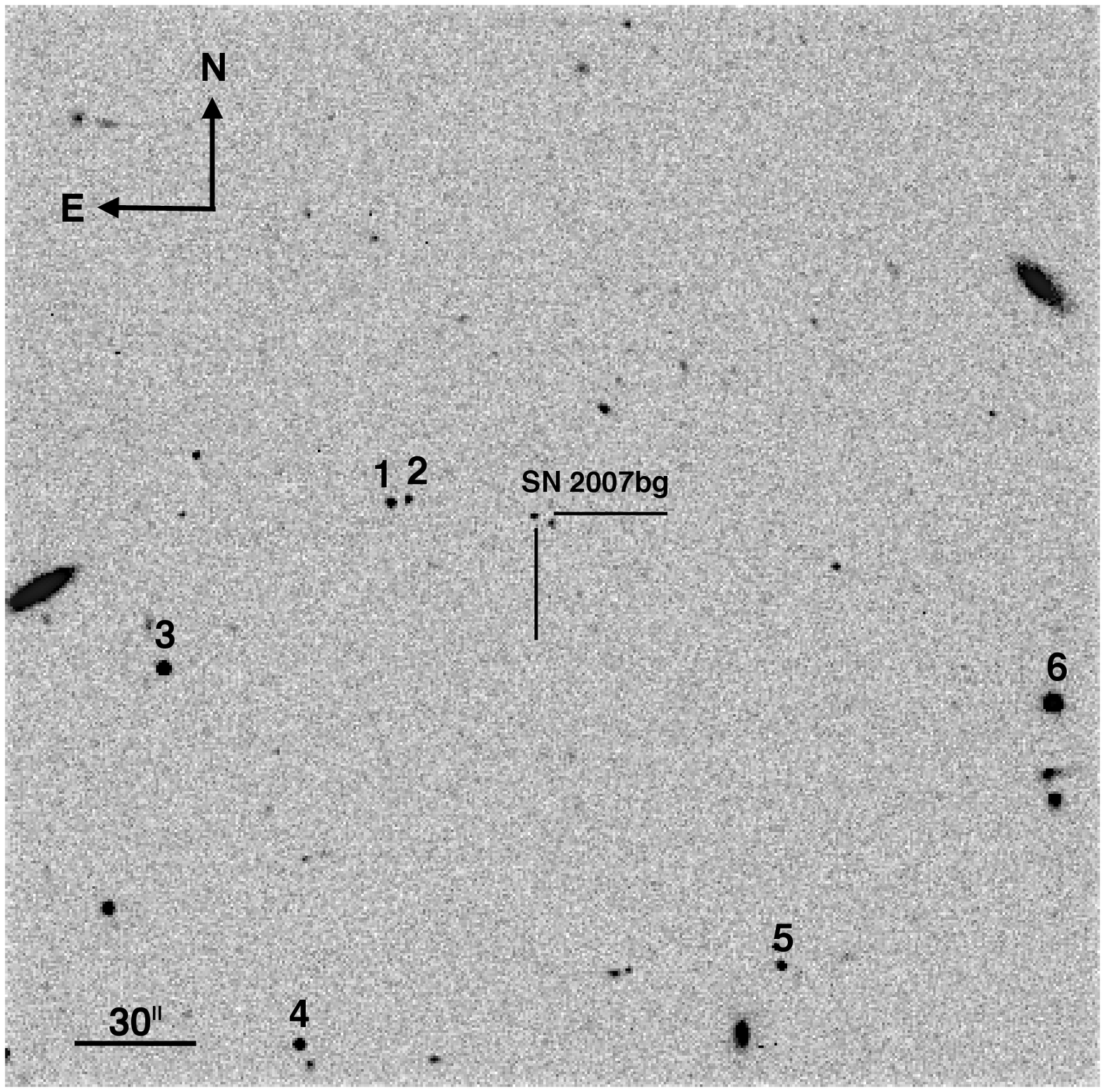}
\caption{The field of SN 2007bg. Local standard stars used to calibrate the photometry of the SN are labelled $1-6$. TNG {\it B}-band image taken on 2007 May 11th, 25 days post-discovery.}
\label{fig:2007bg_sequence_fig}
\end{minipage}
\hspace{0.5cm}
\begin{minipage}[t]{0.48\linewidth}		
\centering
\includegraphics[scale=0.51]{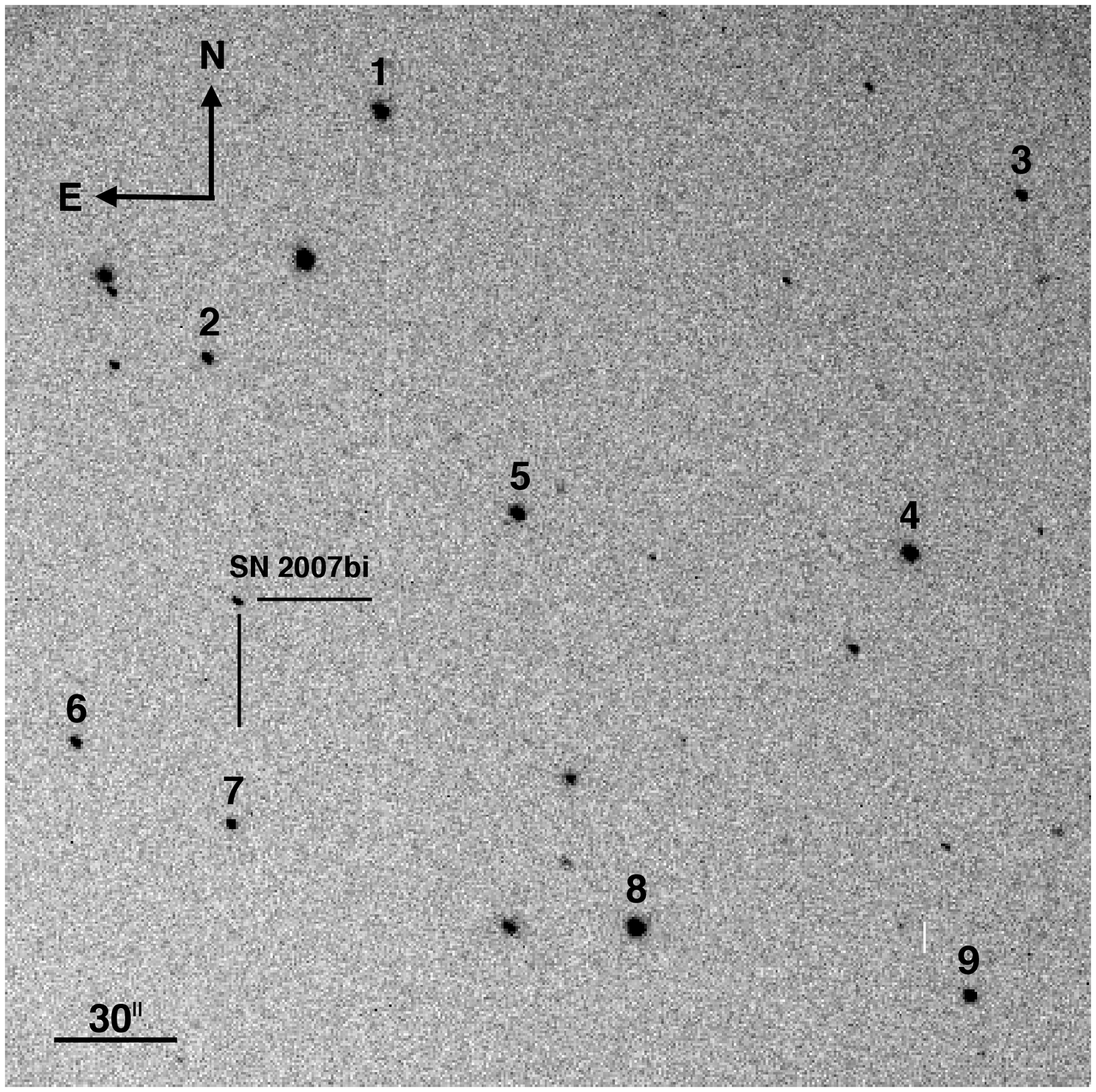}
\caption{The field of SN 2007bi. Local standard stars used to calibrate the photometry of the SN are labelled $1-9$. LT {\it R}-band image taken on 2007 May 12th, 36 days post-discovery. }
\label{fig:2007bi_sequence_fig}
\end{minipage}
\end{figure*}

\cite{Modjaz:2008p87} measured the metallicities at the sites of all GRB associated SNe Ic-BL (GRB/SNe Ic-BL) and compared them to the metallicities measured at the sites of those SNe Ic-BL not found in association with a GRB (non-GRB/SNe Ic-BL). Regardless of the abundance scale employed, \citeauthor{Modjaz:2008p87} found that the environment of every non-GRB/SN Ic-BL was more metal-rich than the environment of any of the GRB/SNe Ic-BL. Upon studying the host galaxies of 42 GRBs discovered within the {\it Hubble} Higher-z Supernova Search, \cite{Fruchter:2006p234} found that GRBs ({\it a}) typically favour low-luminosity, blue, irregular galaxies and ({\it b}) tend to concentrate on the very brightest regions of these host galaxies. This evidence adds weight to the theory that the progenitors of GRBs are massive stars that inhabit low-metallicity environments \citep[e.g. the `Collapsar' model: ][]{Woosley:1993p8292,Woosley:2006p167}. 

With the advent of wide-field, non-targeted SN surveys, more SNe are now being detected in sub-luminous hosts and many of these SNe are proving to be peculiar, e.g. the over-luminous SNe 2005ap \citep{Quimby:2007p50} and 2008es \citep{Gezari:2009p2983,Miller:2009p2870}. Here we report on the follow-up of two such objects discovered by non-targeted SN surveys, both hosted by low-luminosity, dwarf galaxies: SNe 2007bg and 2007bi.

SN 2007bg was discovered in an unfiltered image taken with the 0.45m ROTSE-IIIb telescope \citep[Robotic Optical Transient Search Experiment,][]{Akerlof:2003p26} at the McDonald Observatory, Texas by the Texas Supernova Survey. The object was discovered on 2007 April 16.15 (UT dates are used throughout this paper) located at $\alpha = 11^{h}49^{m}26^{s}.18$, $\delta = +51^{o}49^{'}21^{''}.8$ (J2000.0) \citep{Quimby:2007p3598}. The SN was also recovered in unfiltered CCD images taken by the ROTSE-IIId telescope in Bakirlitepe, Turkey on 2007 April 16.81. No source was recovered at the location of the SN in images taken by the ROTSE-IIIb telescope on 2007 April 6.15, providing an upper-limit to the magnitude of the SN at that epoch. A spectrum taken on 2007 April 18.30 with the 9.2-m Hobby-Eberly Telescope led \citeauthor{Quimby:2007p3598} to classify this object as a Type Ic, claiming similarity with SN 2002ap at around 11 days post maximum brightness. Upon obtaining a spectrum of SN 2007bg with the Telescopio Nazionale Galileo (TNG) on 2007 April 21.04, \citet{Harutyunyan:2007p5835} confirm the resemblance to other SNe Ic-BL. They report that the spectrum is dominated by a broad, strong feature centered at about 6524$\AA$ and state that if this feature is associated with H$\alpha$ at a velocity of $\sim$18 000 km s$^{-1}$ then SN 2007bg should be re-classified as a broad-lined Type II similar to SN 2003bg \citep{Hamuy:2003p6874,Hamuy:2009p12161,Mazzali:2009p12162}.

\begin{table}[t]		
\begin{center}
\caption{\textrm {Local standard stars used to calibrate the photometry of SN 2007bg to the {\it BVRI} standard Landolt system (Johnson-Cousins).}\label{2007bg_sequence}}
\begin{tabular}{c c c c c}
\hline\hline
{Star ID} & {\it B} & {\it V} & {\it R} & {\it I}  \\
\hline
1	& 19.45  {\tiny$\pm$	0.03}		& 19.07  {\tiny$\pm$	0.04}		& 18.72  {\tiny$\pm$ 0.03}		& 18.53  {\tiny$\pm$	0.04}	 \\	
2	& 20.71  {\tiny$\pm$	0.07}		& 19.36  {\tiny$\pm$	0.04}		& 18.42  {\tiny$\pm$ 0.05}		& 17.06  {\tiny$\pm$	0.02}	 \\	
3	& 17.85  {\tiny$\pm$	0.02}		& 17.29  {\tiny$\pm$	0.02}		& 16.87  {\tiny$\pm$ 0.02}		& 16.58  {\tiny$\pm$	0.02}	 \\	
4	& 18.91  {\tiny$\pm$	0.03}		& 18.03  {\tiny$\pm$	0.02}		& 17.34  {\tiny$\pm$ 0.03}		& 16.86  {\tiny$\pm$	0.02}	\\	
5	& 19.68  {\tiny$\pm$	0.04}		& 19.34  {\tiny$\pm$	0.04}		& 19.08  {\tiny$\pm$ 0.04}		& 18.82  {\tiny$\pm$	0.04}	 \\	
6	& 16.00  {\tiny$\pm$	0.02}		& 15.49  {\tiny$\pm$	0.02}		& 15.02  {\tiny$\pm$ 0.02}		& 14.72  {\tiny$\pm$	0.02}	 \\
\hline
\end{tabular}
\end{center}
\end{table}

SN 2007bi was discovered in images taken by the QUEST II camera on the Palomar Oschin 1.2-m Schmidt telescope as part of the NEAT component of the Palomar-QUEST Consortium \citep{Djorgovski:2008p1193} by the Nearby Supernova Factory collaboration \citep{Aldering:2002p3} on 2007 April 6.5 \citep{Gal-Yam:2009}. The object is located at $\alpha = 13^{h}19^{m}20^{s}.19$, $\delta = +08^{o}55^{'}44^{''}.3$ (J2000.0). A spectrum of the object obtained with the Supernova Integral Field Spectrograph on the University of Hawaii 2.2-m telescope under poor conditions on 2007 April 8.6 led \citeauthor{Gal-Yam:2009} to tentatively classify SN 2007bi as a Type Ic. Further spectra taken on 2007 April 15.6 and April 16.4 with the Low Resolution Imaging Spectrometer on the Keck I 10-m telescope, again under poor conditions, helped \citeauthor{Gal-Yam:2009} to confirm the classification of SN 2007bi as a peculiar Type Ic, analogous to SN 1999as \citep{Knop:1999p6878, Hatano:2001p4590} at 25 days post-discovery.

\begin{table}[t]		
\begin{center}
\caption{\textrm {Local standard stars used to calibrate the photometry of SN 2007bi to the {\it BVRI} standard Landolt system (Johnson-Cousins).}\label{2007bi_sequence}}
\begin{tabular}{c c c c c}
\hline\hline
{Star ID} & {\it B} & {\it V} & {\it R} & {\it I}  \\
\hline
1	& 17.91  {\tiny$\pm$	0.02}		& 16.53  {\tiny$\pm$	0.02}		& 15.60  {\tiny$\pm$ 0.02}		& 14.78  {\tiny$\pm$	0.01}	 \\	
2	& 18.27  {\tiny$\pm$	0.02}		& 17.82  {\tiny$\pm$	0.01}		& 17.53  {\tiny$\pm$ 0.02}		& 17.17  {\tiny$\pm$	0.01}	 \\	
3	& 18.38  {\tiny$\pm$	0.02}		& 17.85  {\tiny$\pm$	0.02}		& 17.52  {\tiny$\pm$ 0.02}		& 17.13  {\tiny$\pm$	0.01}	 \\	
4	& 17.15  {\tiny$\pm$	0.01}		& 16.45  {\tiny$\pm$	0.01}		& 15.99  {\tiny$\pm$ 0.01}		& 15.58  {\tiny$\pm$	0.01}	\\	
5	& 17.17  {\tiny$\pm$	0.01}		& 16.65  {\tiny$\pm$	0.01}		& 16.31  {\tiny$\pm$ 0.01}		& 15.95  {\tiny$\pm$	0.01}	 \\	
6	& 18.56  {\tiny$\pm$	0.02}		& 18.00  {\tiny$\pm$	0.02}		& 17.68  {\tiny$\pm$ 0.02}		& 17.32  {\tiny$\pm$	0.01}	 \\	
7	& 19.92  {\tiny$\pm$	0.03}		& 18.42  {\tiny$\pm$	0.02}		& 17.50  {\tiny$\pm$ 0.02}		& 16.56  {\tiny$\pm$	0.01}	 \\	
8	& 17.08  {\tiny$\pm$	0.02}		& 16.05  {\tiny$\pm$	0.01}		& 15.32  {\tiny$\pm$ 0.01}		& 14.78  {\tiny$\pm$	0.01}	 \\	
9	& 19.12  {\tiny$\pm$	0.03}		& 17.81  {\tiny$\pm$	0.02}		& 16.89  {\tiny$\pm$ 0.02}		& 16.18  {\tiny$\pm$	0.01}	 \\	
\hline
\end{tabular}
\end{center}
\end{table}

In this paper we present {\it BVRI} photometric and spectroscopic data sets for SNe 2007bg and 2007bi, discussing techniques of data reduction and our estimations of the distances and extinction towards both SNe in Sections \ref{phot_obs} and \ref{redshifts}. In Sections \ref{lightcurves} and \ref{bol_lcs} we analyse the lightcurves of both SNe, comparing them with other SNe Ic. In Section \ref{spectra} we present spectroscopic data for both SNe, describing the techniques of data reduction. Sections \ref{spec_evol} and \ref{spec_comp} are dedicated to the analysis of the spectra and their comparison with other SNe. In Section \ref{other_para} we present the photospheric velocities, a derivation of the explosion parameters, the search for associated GRBs and an estimation of the oxygen abundances of the host galaxies. Finally we discuss and conclude our findings in Sections \ref{Discussion} and \ref{Conclusions}.

\begin{table*}		
\setlength{\tabcolsep}{5pt}
\begin{minipage}[t]{1.0\textwidth}
\begin{center}
\caption{\textrm {Photometry of SN 2007bg.}\label{2007bg_phot}}
\begin{tabular}{l r r r c r c r c r c r l}
\hline\hline
{UT Date}	& {JD}		& Epoch$^{a}$	& Rest-frame		&{\it B}	& {$K_{BB}$}$^{b}$	&{\it V}	& {$K_{VV}$}$^{b}$	& {\it R}	& {$K_{RR}$}$^{b}$	& {\it I}	& {$K_{II}$}$^{b}$	& {Source}$^{c}$	\\
		& {- 2,450,000}	& (d)			& epoch (d)$^{a}$	&		&  				&		&				&		&				&		&				&				\\
\hline
06/04/07	& 4196.7	& $-$10.0	& $-$9.6	&					&		&					&		&	\multicolumn{1}{l}{(18.20)$^{d}$}	&		&						&		&	R-IIIb$^{e}$ \\
16/04/07	& 4206.7	& 0.0		& 0.0		&					&		&					&		&	\multicolumn{1}{l}{17.7}			& $-$.05	&						&		&	R-IIIb$^{e}$ \\	
16/04/07	& 4207.3	& 0.7		& 0.6		&					&		&					&		&	\multicolumn{1}{l}{17.7}			& $-$.05	&						&		&	R-IIId$^{e}$  \\	
18/04/07	& 4208.7	& 2.0		& 1.9		&					&		&					&		&	\multicolumn{1}{l}{17.8}			& $-$.05	&						&		&	R-IIIb$^{e}$ \\	
23/04/07	& 4214.1	& 7.4		& 7.2		& 19.93 {\tiny $\pm$ .09}	& $-$.18	& 18.98 {\tiny $\pm$ .04}	& $-$.14	&	18.38 {\tiny $\pm$ .05}			& $-$.06	&	18.16 {\tiny $\pm$ .11}	& $-$.01	&	LT \\
27/04/07	& 4218.4	& 11.7	& 11.3			& 20.38 {\tiny $\pm$ .17}	& $-$.18	& 19.35 {\tiny $\pm$ .13}	& $-$.15	&	18.95 {\tiny $\pm$ .09}			& $-$.07	&	18.67 {\tiny $\pm$ .10}	& $-$.01	&	FTN \\
28/04/07	& 4218.9	& 12.2	& 11.9			& 20.48 {\tiny $\pm$ .16}	& $-$.18	& 19.52 {\tiny $\pm$ .07}	& $-$.15	&	18.95 {\tiny $\pm$ .12}			& $-$.07	&	18.69 {\tiny $\pm$ .07}	& $-$.01	&	LT \\	
02/05/07	& 4223.3	& 16.7	& 16.1			& 21.03 {\tiny $\pm$ .41}	& $-$.19	& 19.93 {\tiny $\pm$ .11}	& $-$.16	&	19.35 {\tiny $\pm$ .12}			& $-$.08	&	19.02 {\tiny $\pm$ .09}	& $-$.01	&	FTN \\
05/05/07	& 4225.9	& 19.3	& 18.6			& 21.07 {\tiny $\pm$ .10}	& $-$.19	& 20.09 {\tiny $\pm$ .09}	& $-$.15	&	19.38 {\tiny $\pm$ .07}			& $-$.08	&	19.25 {\tiny $\pm$ .05}	& $-$.01	&	LT \\	
11/05/07	& 4232.6	& 25.4	& 24.5			& 21.24 {\tiny $\pm$ .09}	& $-$.22	& 20.36 {\tiny $\pm$ .11}	& $-$.12	&	19.69 {\tiny $\pm$ .08}			& $-$.08	&	                        & 		&	TNG \\
13/05/07	& 4234.0	& 27.4	& 26.4			& 21.27 {\tiny $\pm$ .09}	& $-$.22	& 20.47 {\tiny $\pm$ .08}	& $-$.12	&	19.72 {\tiny $\pm$ .06}			& $-$.08	&	19.66 {\tiny $\pm$ .12}	& $-$.01	&	LT \\	
19/05/07	& 4240.0	& 33.3	& 32.2			& 21.39 {\tiny $\pm$ .11}	& $-$.21	& 20.54 {\tiny $\pm$ .06}	& $-$.11	&	20.02 {\tiny $\pm$ .07}			& $-$.08	&	                  		& 		&	LT \\	
05/06/07	& 4257.0	& 50.3	& 48.6			& 21.46 {\tiny $\pm$ .06}	& $-$.18	& 20.61 {\tiny $\pm$ .20}	& $-$.09	&									&		&							&		&   LT \\
13/06/07	& 4265.0	& 58.3	& 56.4			& 21.55 {\tiny $\pm$ .09}	& $-$.17	& 20.67 {\tiny $\pm$ .08}	& $-$.07	&	20.32 {\tiny $\pm$ .10}			& $-$.12	&	20.02 {\tiny $\pm$ .19}	& .00	&	LT \\	
07/07/07	& 4289.0	& 82.3	& 79.6			& 21.67 {\tiny $\pm$ .07}	& $-$.13	& 20.93 {\tiny $\pm$ .08}	& $-$.02	&	20.74 {\tiny $\pm$ .08}			& $-$.15	&	20.47 {\tiny $\pm$ .14}	& .00	&	LT \\
12/08/07	& 4324.9	& 117.4	& 113.4			& 22.12 {\tiny $\pm$ .08}	& $-$.07	& 21.61 {\tiny $\pm$ .09}	& .05	&	21.32 {\tiny $\pm$ .13}			& $-$.20	&	20.95 {\tiny $\pm$ .13}	& .00	&	WHT \\
\hline
\end{tabular}
\end{center}
{\it Notes.} Errors on all magnitudes have been determined by taking into account the uncertainty of PSF fitting of the SN, the uncertainty of having accurately removed the background contamination measured via artificial-star experiments and the uncertainty of both calculating the colour-terms and zero-points of each night's imaging via the measurement of a local standard stars in the field of the SN. \\
$^{a}$ Relative to the date of discovery (CBET 927) \\
$^{b}$ {\it K}-corrections determined for all epochs with spectral information and extrapolated to all other epochs (decimals). \\
$^{c}$ R-IIIb =  0.45m ROTSE-IIIb,  McDonald Observatory, Texas (USA). R-IIId =  0.45m ROTSE-IIId, Turkish National Observatory, Bakirlitepe (Turkey). LT = RATCam on the 2.0m Liverpool Telescope, Observatorio del Roque de Los Muchachos, La Palma (Spain). FTN = HawkCam on the 2.0m Faulkes Telescope North, Haleakala Observatory, Hawaii (USA). TNG = DOLoRes on the 3.58m Telescopio Nazionale Galileo, Observatorio del Roque de Los Muchachos, La Palma (Spain). WHT = auxiliary port camera on the 4.2m William Herschel Telescope, Observatorio del Roque de Los Muchachos, La Palma (Spain).\\
$^{d}$ Upper magnitude limit taken from CBET 927.\\
$^{e}$ Magnitude taken from CBET 927.
\end{minipage}
\end{table*}

\begin{table*}		
\setlength{\tabcolsep}{5pt}
\begin{minipage}[t]{1.0\textwidth}
\begin{center}
\caption{\textrm {Photometry of SN 2007bi.}\label{2007bi_phot}}
\begin{tabular}{l r r r l r l r l r l r l}
\hline\hline
{UT Date}	& {JD}		& Epoch$^{a}$	& Rest-frame		&{\it B}	& {$K_{BB}$}$^{b}$	&{\it V}	& {$K_{VV}$}$^{b}$	& {\it R}	& {$K_{RR}$}$^{b}$	& {\it I}	& {$K_{II}$}$^{b}$	& {Source}$^{c}$	\\
		& {- 2,450,000}	& (d)			& epoch (d)$^{a}$	&		&  				&		&				&		&				&		&				&				\\
\hline
21/02/07 & 4153.0	& 0.0		& 0.0		&								&		&								&		& \multicolumn{1}{l}{17.64}	& 				& 						&		& G09$^{d}$ \\
06/04/07 & 4197.0	& 44.00		& 39.04		& 								&		& 								&		& \multicolumn{1}{l}{18.3}		& .00		& 						&		& POS$^{e}$ \\
22/04/07 & 4213.6	& 60.59		& 53.76		& 18.83 {\tiny $\pm$ .06}		& $-$.36	& 18.50 {\tiny $\pm$ .10}		& $-$.09	& 18.27 {\tiny $\pm$ .04}		& $-$.03	& 18.01 {\tiny $\pm$ .07}		& $-$.20	& Ekar \\
23/04/07 & 4214.0	& 61.02		& 54.14		& 18.86 {\tiny $\pm$ .03}		& $-$.35	& 18.45 {\tiny $\pm$ .03}		& $-$.10	& 18.34 {\tiny $\pm$ .02}		& $-$.04	& 18.03 {\tiny $\pm$ .05}		& $-$.16	& LT\\
02/05/07 & 4223.3	& 70.34		& 62.41		& 19.07 {\tiny $\pm$ .07}		& $-$.32	& 18.60 {\tiny $\pm$ .04}		& $-$.12	& 18.35 {\tiny $\pm$ .02}		& $-$.08	& 17.97 {\tiny $\pm$ .04}		& $-$.10	& FTN \\
05/05/07 & 4226.0	& 72.96		& 64.74		& 19.03 {\tiny $\pm$ .03}		& $-$.32	& 18.65 {\tiny $\pm$ .03}		& $-$.12	& 18.41 {\tiny $\pm$ .04}		& $-$.08	& 17.96 {\tiny $\pm$ .12}		& $-$.12	& LT \\
12/05/07 & 4233.0	& 79.97		& 70.96		& 19.20 {\tiny $\pm$ .02}		& $-$.31	& 18.69 {\tiny $\pm$ .02}		& $-$.11	& 18.51 {\tiny $\pm$ .02}		& $-$.09	& 18.10 {\tiny $\pm$ .05}		& $-$.15	& LT \\
18/05/07 & 4239.3	& 86.31		& 76.58		& 19.23 {\tiny $\pm$ .04}		& $-$.30	& 18.76 {\tiny $\pm$ .03}		& $-$.11	& 18.54 {\tiny $\pm$ .04}		& $-$.10	& 18.20 {\tiny $\pm$ .07}		& $-$.19	& FTN \\
19/05/07 & 4234.0	& 86.94		& 77.14		& 19.27 {\tiny $\pm$ .03}		& $-$.30	& 18.83 {\tiny $\pm$ .03}		& $-$.11	& 18.57 {\tiny $\pm$ .03}		& $-$.10	& 		       				& 		& LT \\
31/05/07 & 4251.9	& 98.92		& 87.77		& 19.67 {\tiny $\pm$ .17}		& $-$.29	& 18.96 {\tiny $\pm$ .10}		& $-$.10	& 18.86 {\tiny $\pm$ .14}		& $-$.11	& 18.48 {\tiny $\pm$ .12}		& $-$.25	& LT \\
07/07/07 & 4288.9	& 135.92	& 120.60	& 20.43 {\tiny $\pm$ .02}		& $-$.32	& 19.68 {\tiny $\pm$ .03}		& $-$.10	&     						&		& 18.80 {\tiny $\pm$ .09}		& $-$.15	& LT \\
12/08/07 & 4324.0	& 171.88	& 152.51	& 20.73 {\tiny $\pm$ .07}		& $-$.37	& 19.94 {\tiny $\pm$ .12}		& $-$.11	& 19.64 {\tiny $\pm$ .08}		& $-$.16	& 19.06 {\tiny $\pm$ .11}		& $-$.04	& WHT \\ 
10/04/08 & 4566.2	& 413.19	& 366.63	& $>$22.59$^{f}$				& $-$.54	& 22.82 {\tiny $\pm$ .24}		& $-$.18	& 22.23 {\tiny $\pm$ .22}		& $-$.07	& 21.89 {\tiny $\pm$ .07}		& $-$.08	& VLT \\
29/04/09 & 4951.1	& 754.10	& 669.12	& $\gtrsim$22.7					&			& $\gtrsim$22.9					&			& $\gtrsim$22.6												&								& $\gtrsim$22.2												&								& LT \\
\hline
\end{tabular}
\end{center}
{\it Notes.} Errors on all magnitudes have been determined by taking into account the uncertainty of PSF fitting of the SN, the uncertainty of having accurately removed the background contamination measured via artificial-star experiments and the uncertainty of both calculating the colour-terms and zero-points of each night's imaging via the measurement of a local standard stars in the field of the SN. \\
$^{a}$ Relative to the date of maximum brightness \citep{Gal-Yam:2009}. \\
$^{b}$ {\it K}-corrections determined for all epochs with spectral information and extrapolated to all other epochs (decimals). \\
$^{c}$ POS = QUEST II camera on the Palomar Oschin 1.2m Schmidt telescope, Palomar Observatory, Califonia (USA).  Ekar = AFSOC on the 1.82m Copernico Telescope, INAF, Osservatorio di Asiago, Mt. Ekar, Asiago (Italy). LT = RATCam on the 2.0m Liverpool Telescope, Observatorio del Roque de Los Muchachos, La Palma (Spain). FTN = HawkCam on the 2.0m Faulkes Telescope North, Haleakala Observatory, Hawaii (USA). TNG = DOLoRes on the 3.58m Telescopio Nazionale Galileo, Observatorio del Roque de Los Muchachos, La Palma (Spain). WHT = auxiliary port camera on the 4.2m William Herschel Telescope, Observatorio del Roque de Los Muchachos, La Palma (Spain). VLT = FORS2 on the 8.2m Very Large Telescope (unit telescope 1), Paranal Observatory, Cerro Paranal in the Atacama Desert (Chile).\\
$^{d}$ Peak magnitude determined by \cite{Gal-Yam:2009} via parabolic fitting to pre- and post-peak magnitude data.
$^{e}$ Magnitude taken from CBET 929.
$^{f}$ Upper-limit: flux dominated by host-galaxy.
\end{minipage}
\end{table*}

\section{Photometry}\label{obs_phot}

\subsection{Photometric Observations and Data Reduction}\label{phot_obs} 

Images of SNe 2007bg and 2007bi were obtained during a monitoring campaign which commenced on JD 245\,413.6 ($\sim$7 d and $\sim$15 d post-discovery of SNe 2007bg and 2007bi respectively) and lasted for approximately 110 d. The campaign mainly involved RATCam on the 2.0-m Liverpool Telescope (LT) and HawkCam on the 2.0-m Faulkes Telescope North (FTN). Images where also obtained with AFOSC on the 1.82-m Copernico Telescope, DOLoRes on the 3.58-m Telescopio Nazionale Galileo (TNG) and the Auxiliary Port Imager (AUX) on the 4.2-m William Herschel Telescope (WHT). Our photometric data set of SN 2007bi also includes late-time images (369 d post-discovery) taken with the FORS2 instrument on the Very Large Telescope (VLT).

Apart from the LT images which were received pre-reduced by the LT pipeline, all images were reduced (overscan corrected, bias-subtracted and flat-fielded) using standard {\tiny IRAF}\footnote{IRAF is distributed by the National Optical Astronomy Observatories, which are operated by the Association of the Universities for Research in Astronomy, Inc., under contract to the National Science Foundation.} routines. LT observations of Landolt standard-stars \citep{Landolt:1992p2508} taken on the photometric nights of 2007 May 12 and May 13 were used to calibrate local sequences of stars in the fields of view of SNe 2007bi and 2007bg respectively (see Figures \ref{fig:2007bg_sequence_fig} and \ref{fig:2007bi_sequence_fig}). The magnitudes of these local standard stars, presented in Tables \ref{2007bg_sequence} and \ref{2007bi_sequence}, were used to calculate the photometric zero-points and colour-terms required to determine the relative magnitudes of the SNe in the other images included in our data set. We used the SNOoPY\footnote{SNOoPY, originally presented by \cite{Patat:1996p4219}, has been implemented in IRAF by E. Cappellaro. The package is based on DAOPHOT, but optimised for SN magnitude measurements.} package implemented in {\tiny IRAF} to measure the average point-spread function (PSF) of each image from suitable field stars. This PSF was then used to fit the profile of the SN while simultaneously fitting a polynomial surface to, and removing the background from the region of the SN. Artificial stars of the same magnitude and profile as the SN were generated near to the location of the SN and PSF fitted in the background subtracted image. The deviations in the measured magnitudes of these recovered artificial stars were used to estimate errors in the SN photometry. 

In the late-time VLT images of SN 2007bi the flux of the SN has declined to the point where it is comparable to the flux of its underlying host galaxy. In order to determine an accurate measure of the magnitude of the SN it is therefore vital to remove the contaminating flux of the host galaxy. This is achieved by taking very late-time {\it template} images of the location of the SN (when we can be sure that the SN has faded to the point where it contributes a negligible fraction to flux at its location) and subtracting these template images from the {\it target} images which include the SN. If the image subtraction is clean then the residual flux at the location of the SN should belong purely to the SN. {\it BVr'i'} template images of the location of SN 2007bi were obtained with the LT on 2009 April 29, $\sim$754 d post-discovery. Once we had geometrically registered the template images to the target images, we fitted PSFs to a selection of sources common to both images. The {\it mrj$\_$phot} routine within the {\it ISIS} package \citep{Alard:2000p9126} was used to compare these sources and derive an optimal convolution kernel used to match the two images so that they could be differenced. Finally, the residual flux at the location of the SN was measured using the SNOoPY package. Unfortunately the {\it B}-band template image was not deep enough to allow us to subtract out the host-galaxy flux from the VLT {\it B}-band image; therefore we include this data point as an upper-limit to the supernova magnitude at this epoch. In the future we hope to obtain deeper, late-time images of the location of SN 2007bi to further constain the late-time photometry of the object via template subtraction.

\begin{figure*}		
\begin{minipage}[t]{0.48\linewidth}
\centering
\includegraphics[scale=0.52,angle=270]{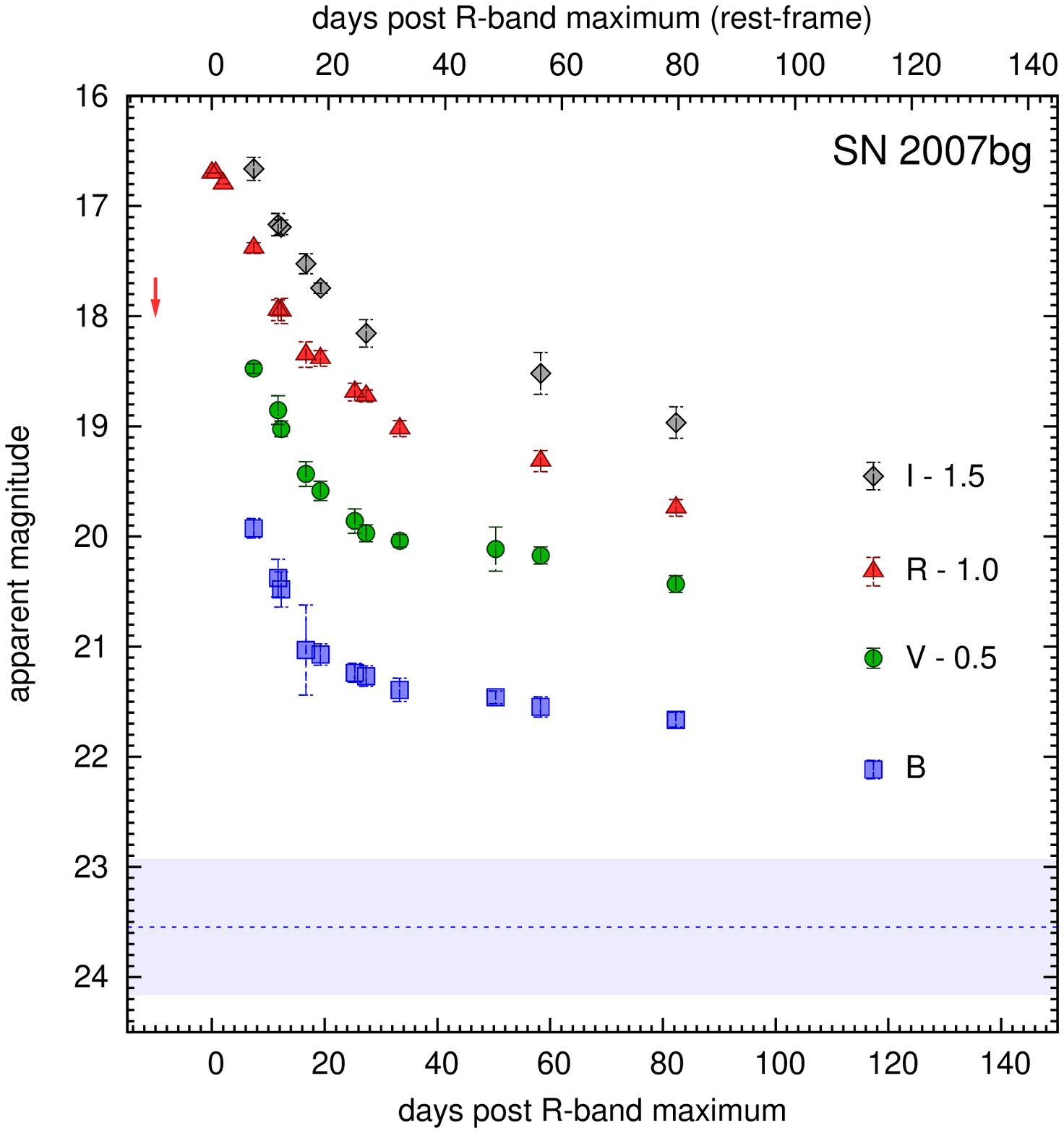}
\caption{{\it BVRI} light curves for SN 2007bg plotted relative to the date of {\it R}-band maximum light. Magnitudes have not been corrected for extinction and {\it K}-corrections have not been applied. A pre-discovery limiting-magnitude reported in CBET 927 provides an upper-limit to the $R$-band magnitude (red arrow). The blue dashed line (shaded area) indicates the {\it B}-band magnitude (magnitude error) of the host galaxy as measured from pre-explosion SDSS images (see Section \ref{hosts}).}
\label{07bg_lc}
\end{minipage}
\hspace{0.5cm}
\begin{minipage}[t]{0.48\linewidth}
\centering
\includegraphics[scale=0.52, angle=270]{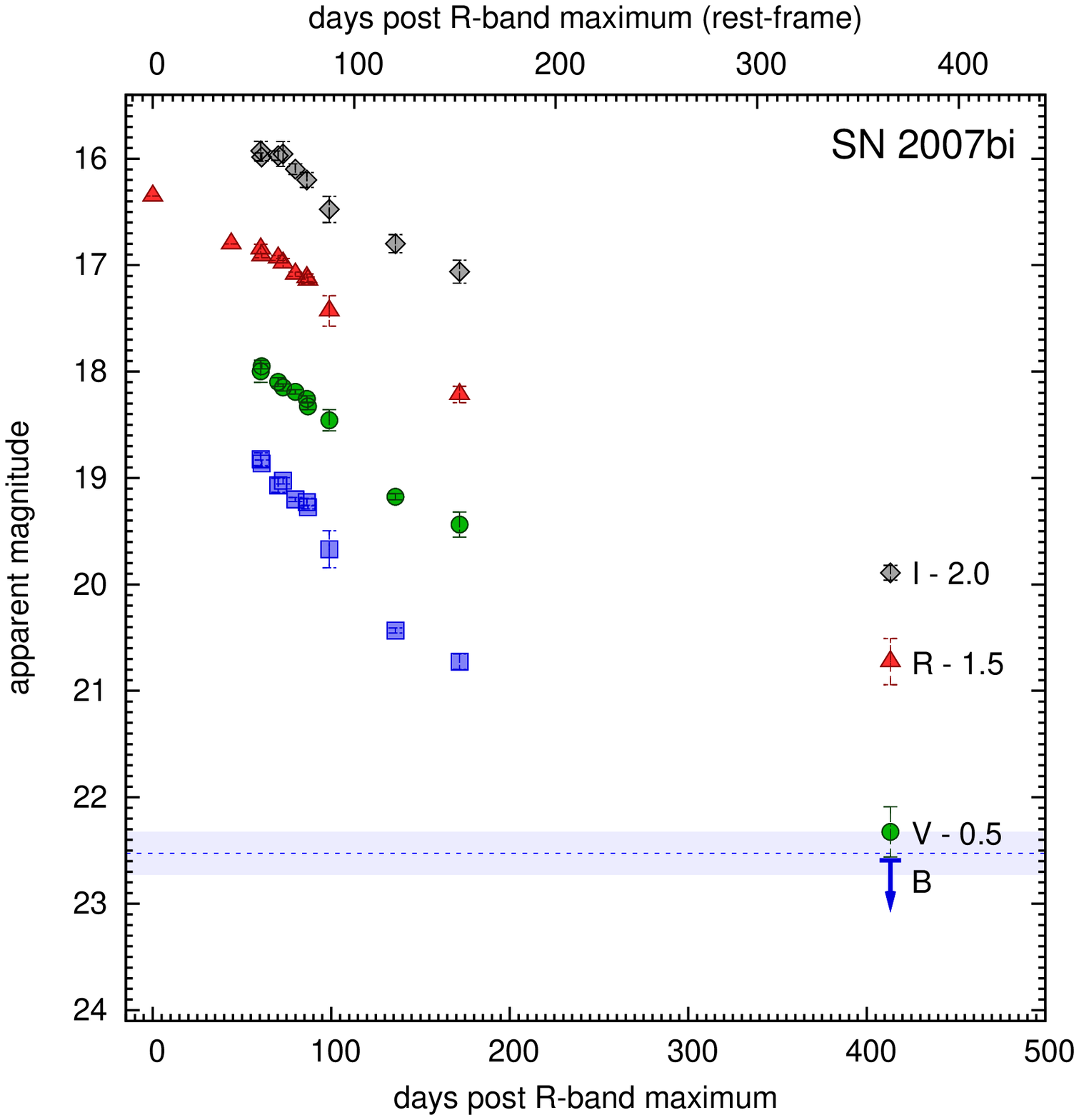}
\caption{{\it BVRI} light curves for SN 2007bi plotted relative to the date of {\it R}-band maximum light. Magnitudes have not been corrected for extinction and {\it K}-corrections have not been applied. Peak {\it R}-band magnitude from \cite{Gal-Yam:2009}. The blue dashed line (shaded area) indicates the {\it B}-band magnitude (magnitude error) of the host galaxy as measured from pre-explosion SDSS images (see Section \ref{hosts}). The final point in the {\it B}-band light represents an upper-limit.}
\label{07bi_lc}
\end{minipage}
\end{figure*}

At respective redshifts of 0.0347 and 0.127 (see Section \ref{redshifts}), the {\it K}-corrections for SNe 2007bg and 2007bi are significant. Making use of our spectral data sets, we determined the differences between synthetic photometry of the SNe as measured in the observed and rest frames of reference, i.e. the {\it K}-corrections. Corrections between 0.00 and 0.22 mags were measured for SN 2007bg and between 0.00 and 0.60 mags for SN 2007bi, depending both on the wavelength-band and the phase of the SN evolution. {\it BVRI} photometric data for SNe 2007bg and 2007bi can be found in Tables \ref{2007bg_phot} and \ref{2007bi_phot} and observed light curves in Figures \ref{07bg_lc} and \ref{07bi_lc} respectively. 

\subsection{Distance and Extinction}\label{redshifts}

In the latest spectra taken of SNe 2007bg and 2007bi, narrow emission-lines from the host galaxies can be clearly distinguished from broad spectral features of the SNe. These lines (including H$\alpha$, H$\beta$, [O {\tiny II}] $\lambda3727$ and [O {\tiny III}] ${\rm \lambda\lambda}4959,5007$) are used to determine redshifts of $z=0.0346$ and $z=0.127$ to SNe 2007bg and 2007bi respectively.  Assuming $H_o$ = 70 km s$^{-1}$ Mpc$^{-1}$ these redshifts equate to distance moduli of $\mu$=35.84 and $\mu$=38.85. None of the spectra show detection of a narrow Na {\tiny I} D absorption line associated with either of the host galaxies. As there is a general correlation between the strength of this Na {\tiny I} D line and the colour excess ascribed to a galaxy \citep{Turatto:2003p9889}, we assume both SNe suffer from little, if any, host-galaxy extinction. Within the uncertainities, the H$\alpha$/H$\beta$ Balmer decrements measured for the host galaxies is close to the theorectical value of 2.86 \citep[assuming case B recombination,][]{Osterbrock:1989p8676}, further adding weight to the assumption that these galaxies are subject to little intrinsic reddening. This is to be expected as both SNe are hosted by sub-luminous, blue, dwarf galaxies of presumably low-metallicity (see Section \ref{hosts}) and low dust content. We therefore assign only a Galactic component \citep{Schlegel:1998p99} to the colour excess of each SN: $E(B-V) = 0.021$ mag for SN 2007bg and $E(B-V) = 0.028$ mag for SN 2007bi.

\subsection{Light Curves}\label{lightcurves}

\begin{figure*}		
\begin{minipage}[t]{1.0\linewidth}		
\centering
\includegraphics[scale=0.7]{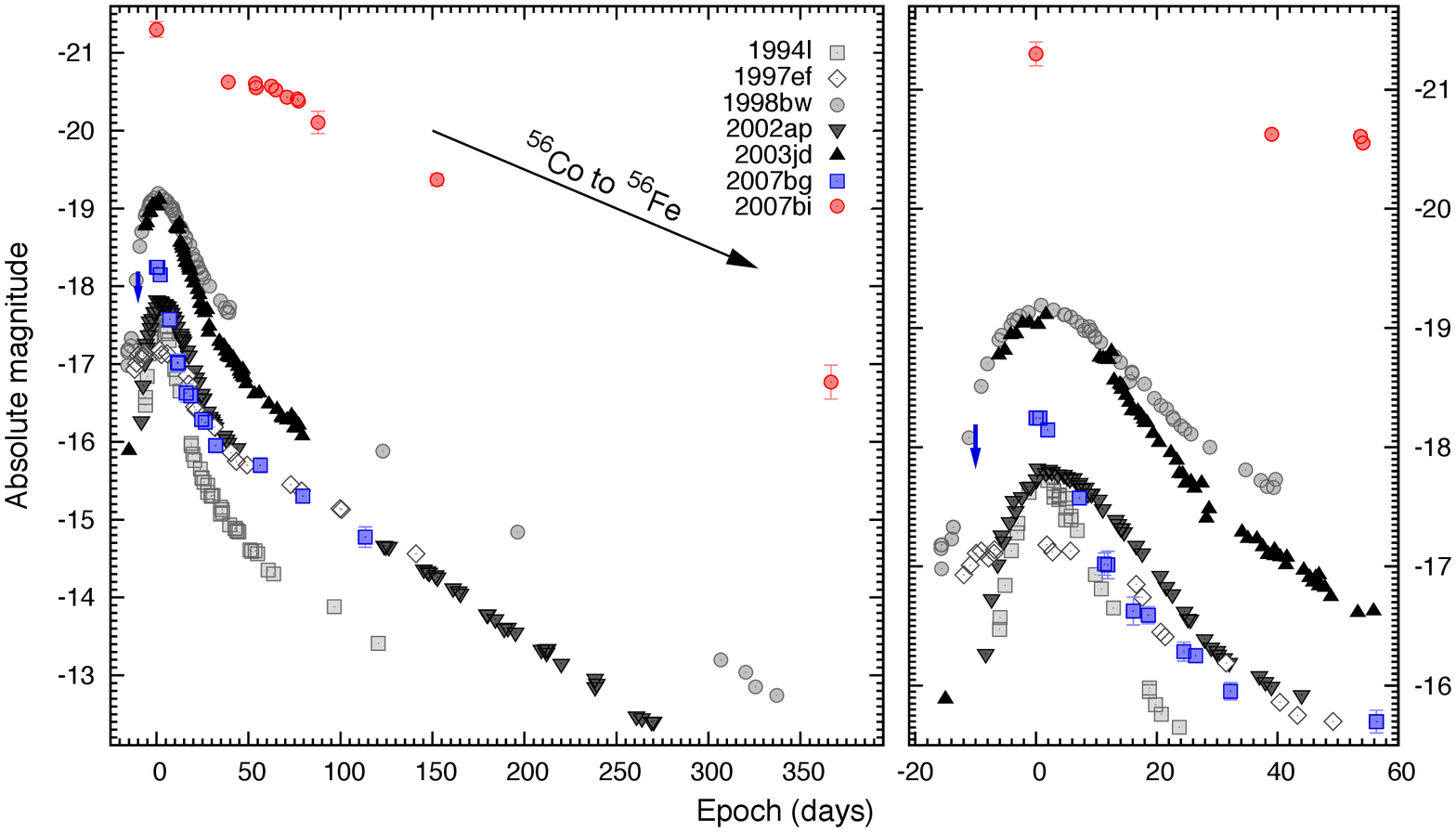}
\caption{\textrm{{\it R}-band lightcurve comparisons of SNe Ic ({\it V}-band used for SN1997ef comparison). All comparison data have been corrected for extinction and epochs are relative to the {\it R}-band maximum light as stated in the literature and given in the rest frame of the individual SNe. {\it References and colour-excess:} SN 1994I with $E(B-V)=0.300$ \citep{Richmond:1996p4690}, SN 1997ef with $E(B-V)=0.042$ \citep{Iwamoto:1998p4660}, 1998bw with $E(B-V)=0.065$ \citep{Galama:1998p4661,McKenzie:1999p5100,Sollerman:2000p5093,Patat:2001p4674}, SN 2002ap with $E(B-V)=0.09$ \citep{Foley:2003p136, Yoshii:2003p5199, Pandey:2003p5268, Tomita:2006p5215}, SN 2003jd with $E(B-V)=0.14$ \citep{Valenti:2008p66}. The panel on the right provides an expanded view of the early-time behaviour of the lightcurves.}
\label{RLC-comp-plot}}
\end{minipage}
\begin{minipage}[t]{1.0\linewidth}		
\centering
\includegraphics[scale=0.7]{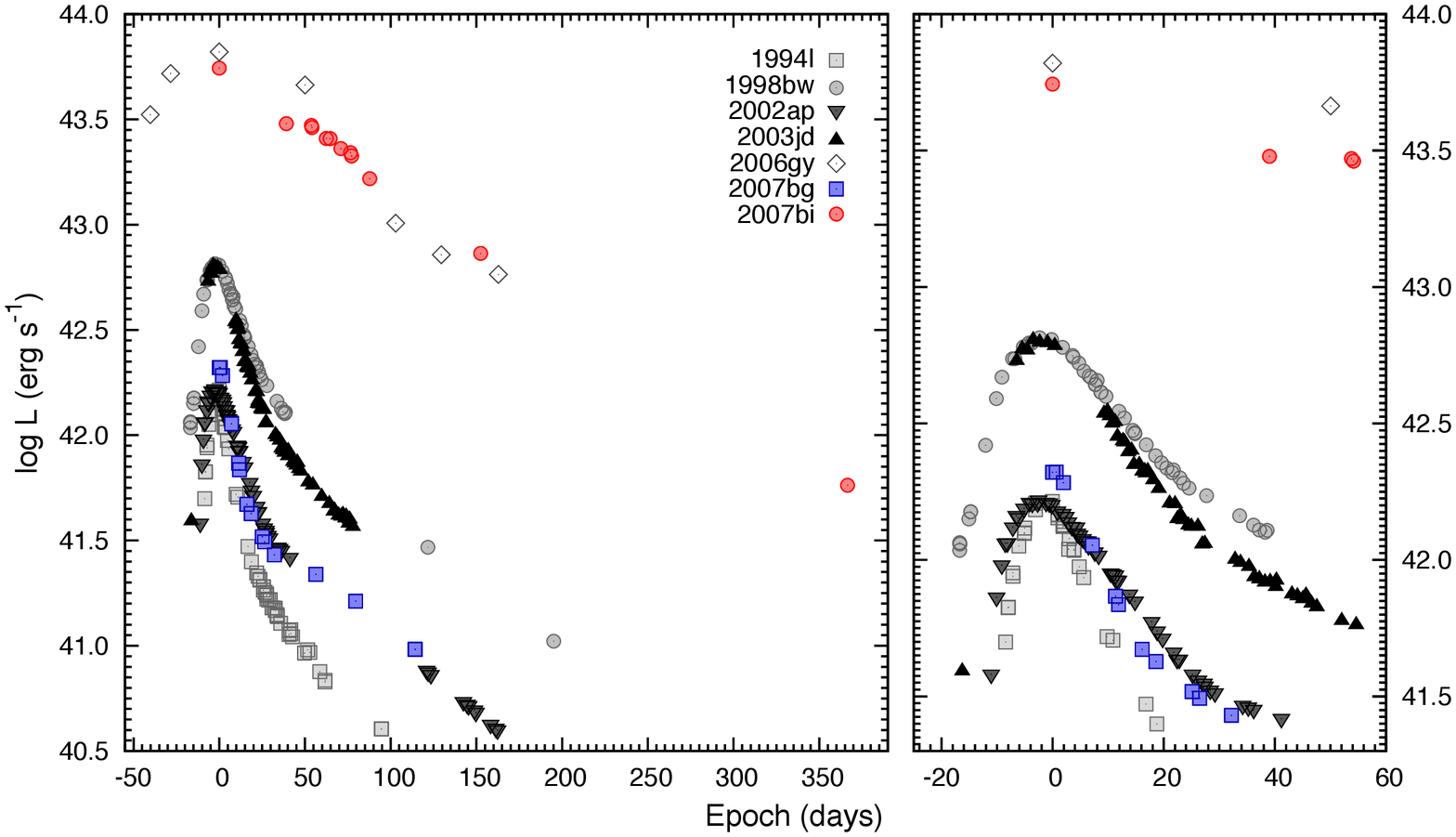}
\caption{\textrm{{\it BVRI} quasi-bolometric lightcurve comparisons. All comparison data have been corrected for extinction and epochs are relative to the {\it R}-band maximum light and are give in the rest-frame of the individual SNe. {\it References and colour-excess:} As in Figure \ref{RLC-comp-plot}, also SN 2006gy with $E(B-V)=0.56$ \citep{Agnoletto:2009p3196}. The panel on the right provides an expanded view of the early-time behaviour of the lightcurves.}
\label{bolo-comp-plot}}
\end{minipage}
\end{figure*}

\begin{table}		
\begin{center}
\caption{\textrm {Summary information for SNe 2007bg and 2007bi}\label{summary_table}}
\begin{tabular}{l l l}
\hline\hline
{} & SN 2007bg & SN 2007bi \\
\hline
RAJ2000						& $11^{h}49^{m}26{s}.18$		& $13^{h}19^{m}20{s}.19$	\\	
DECJ2000					& $+51^{o}49^{'}21^{''}.8$		& $+08^{o}55^{'}44^{''}.3$		\\
Discovery date				& 2007 April 16					& 2007 April 6			\\
Julian date					& 2,454,206.7					& 2,454,197.0				\\
{\it z}						& 0.0346						& 0.127					\\
Maximum brightness (M$_R$)	& $-$18.2						& $-21.3$ {\tiny$\pm0.1$}$^a$ \\
Host Galaxy					& Anon.							& SDSS		\\
									&						& J131920.14+085543.7		\\
Host galaxy magnitude (M$_B$) 	& $-12.4$ {\tiny$\pm0.6$}		& $-16.4$ {\tiny $\pm0.2$} \\
Host galaxy H$\alpha$/H$\beta$			& 2.59 {\tiny $\pm$ 0.39}	& 2.79 {\tiny $\pm$ 1.26} \\
12+log(O/H) PP04$^b$			& $8.18$ {\tiny$\pm0.17$}		& $-$ \\
12+log(O/H) M91$^c$			& $-$						& $8.15$ {\tiny$\pm0.13$} \\
Discovery reference			& 1							& 2 \\
Other references				& 3, 4, 5, 6 					& 7, 8 \\
\hline
\end{tabular}
\end{center}
{\it a.} \cite{Gal-Yam:2009} \\
{\it b.} \cite{Pettini:2004p217} abundance scale\\
{\it c.} \cite{McGaugh:1991p7985} abundance scale.\\
{\it References}: 1.\cite{Quimby:2007p3598}, 2. \cite{Antilogus:2007p5845}, 3. \cite{Harutyunyan:2007p5835}, 4. \cite{Moretti:2007p3644}, 5. \cite{Prieto:2007p5838}, 6. \cite{Soderberg:2007p5839}, 7. \cite{Moretti:2007p3644}, 8.  \cite{Nugent:2007p3645}
\end{table}

The {\it BVRI} light curves of SNe 2007bg and 2007bi cover approximately the first four months of evolution before solar conjunction and are shown in Figures \ref{07bg_lc} and \ref{07bi_lc} respectively. SN 2007bi was also re-observed when it became visible again at +369 days post-discovery. In the {\it R}-band lightcurve of SN 2007bg we include the early-time unfiltered data-points (calibrated against USNO-B1.0 {\it R}-band values) and the upper-limit measured at the location of the SN on 2007 April 6.15, $\sim$10 days before discovery, by the ROTSE telescopes \citep{Quimby:2007p3598}. Constrained by this early-time limit to the magnitude, it is clear that SN 2007bg was discovered close to maximum brightness in the {\it R}-band. Upon recovering SN 2007bi in unfiltered images taken by the Catalina Sky Survey (CSS) \cite{Gal-Yam:2009} have found that the SN peaked approximately 44 d before discovery on JD2454153 $\pm 1$. We include the peak magnitude measured by \cite{Gal-Yam:2009} in our {\it R}-band light curve.

As both SNe 2007bg and 2007bi are located at relatively large distances it is necessary to present these objects within their rest-frames to enable an accurate comparison with other SNe. Calibration of the light curves is performed by making use of the {\it K}-corrections derived (Section \ref{obs_phot}) and correcting for the `stretching' of the light curves induced by time-dilation at these moderate redshifts. Comparison of the absolute-magnitude light curves of SNe 2007bg (blue data-points) and 2007bi (red data-points) with other well-sampled SNe Ic is displayed in Figure \ref{RLC-comp-plot}. All data are plotted in the {\it R}-band except for SN 1997ef which is plotted in the {\it V}-band due to a lack of coverage in the {\it R}-band. As previously mentioned, SNe Ic are known to be a heterogeneous group both in terms of their absolute peak-magnitude range and their lightcurve shapes \citep[see][]{Richardson:2006p72}, not to mention their spectral properties (see Section \ref{spectra}). Comparison SNe Ic presented in this paper have been chosen to span this wide range of properties from the prototypical SN 1994I \citep{Sauer:2006p4972, Richmond:1996p4690} to the very energetic SN 1998bw \citep{Galama:1998p4661, Patat:2001p4674} and including intermediate SNe 2007gr \citep{Valenti:2008p233,Hunter:2009p13049}, 1997ef \citep{Mazzali:2000p4496, Iwamoto:1998p4660, Nakano:1997p5156}, 2002ap \citep{Mazzali:2002p7663,GalYam:2002p5310,Foley:2003p136,Yoshii:2003p5199,Hendry:2005p220}, 2003jd \citep{Valenti:2008p66} and 2004aw \citep{Taubenberger:2006p4726}. In both Figures \ref{RLC-comp-plot} and \ref{bolo-comp-plot} all data are plotted relative to {\it R}-band maximum-light and all light curves have been corrected for time-dilation, i.e. SNe are plotted in their rest-frame. Whenever a recessional velocity (corrected for the Local Group infall into the Virgo Cluster) has been used to determine a kinematic distance modulus to a SN then we have adopted $H_o= 70$ km s$^{-1}$ Mpc$^{-1}$. Extinction values attributed to each SN have been taken from the sources of the photometric data, except for SN 1994I where we make use of the colour excess measurement of  \cite{Sauer:2006p4972}, and can be found in the caption of Figure \ref{RLC-comp-plot}.

\begin{figure}[ht]		
\begin{center}
\includegraphics[scale=0.7, angle=270]{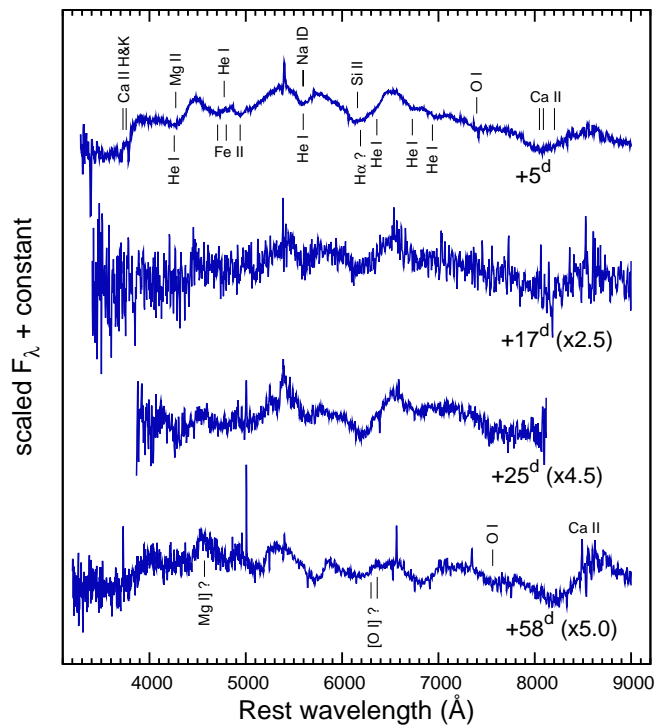}
\caption{{\it Top:} Spectroscopic evolution of SN 2007bg from +5 d to +58 d relative to the {\it R}-band maximum light (2007 April 16). Both wavelength and epochs are stated in the SN rest-frame. Relative scaling factors in parenthesis.}
\label{fig:SN07bg_spectra}
\end{center}
\end{figure}
\begin{figure}[ht]		
\begin{center}
\includegraphics[scale=0.7, angle=270]{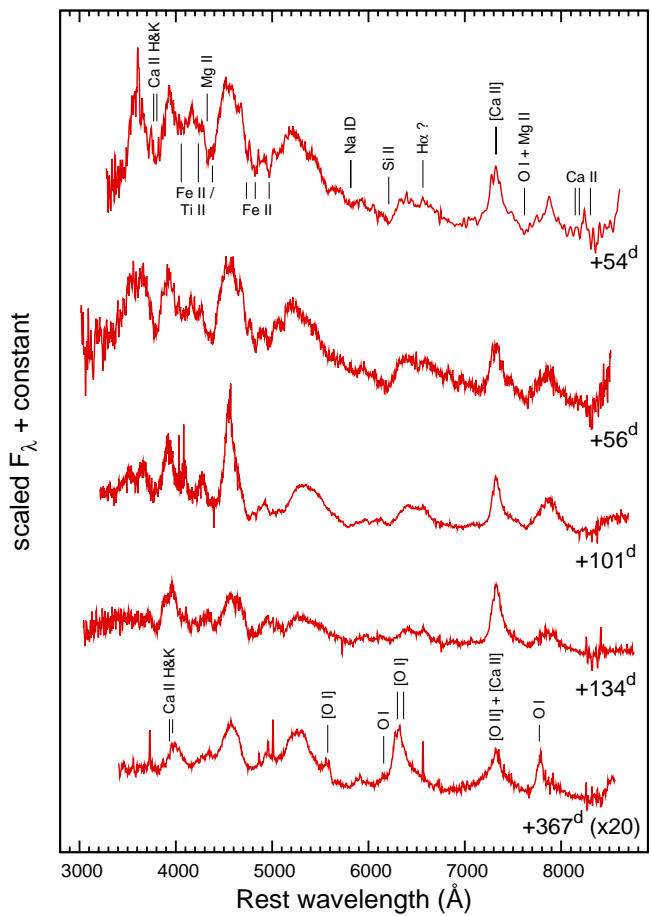}
\caption{Spectroscopic evolution of SN 2007bi from +54 d to +367 d relative to the date of {\it R}-band maximum light \citep[2007 February 21, ][]{Gal-Yam:2009}. Both wavelength and epochs are stated in the SN rest-frame.}
\label{fig:SN07bi_spectra}
\end{center}
\end{figure}

Reaching a peak magnitude of M$_{R}=-18.2$ mag, SN 2007bg is found to be considerably brighter than the prototypical SN 1994I, yet still $\sim$1 mag fainter than luminous SNe 1998bw and 2003jd. The rise to peak brightness of SN 2007bg, constrained by the limiting-magnitude $\sim$10 days before discovery, seems to be similar in speed to that of SN 2002ap and possibly as fast as SN 1994I. SNe 1997ef, 1998bw and 2003jd display much slower rise-times. From maximum brightness to $\sim$30 days, the decline rate of SN 2007bg is remarkably similar to that of SN 1994I. It then mirrors SN 2002ap and begins to decline more slowly at a rate similar to the other SNe Ic-BL 1998bw and 1997ef. This slowing of the decline marks the onset of the radioactive tail. A rapid decline after maximum light followed by a transition to a slower decline rate is typical of all SNe, bar type IIP and IIn. 

The photometric evolution of SN 2007bi is characterised by its remarkably slow post-maximum decline and its extreme peak luminosity. This prolonged decline is consistent across all  {\it BVRI}-bands of the SN 2007bi light curves (see Figure \ref{07bi_lc}). When compared to other SNe Ic, SN 2007bi declines much slower than any other object, its nearest rivals being SNe 1997ef, 1997dq and 1998bw which exhibited little, if any, $\gamma$-ray escape in their late-time light curves \citep{Mazzali:2004p5906}. This extremely slow decline rate of the light curve of SN 2007bi, presumably accompanied by a remarkably broad peak, is indicative of a very large ejected mass. Reaching a peak magnitude M$_{R}\sim-21.3$ mag \citep{Gal-Yam:2009}, SN 2007bi is probably the most luminous SN Ic ever discovered. SN 1999as has been reported to have reached a peak magnitude M$_{V} \gtrsim$ $-$21.5 \citep{Deng:2001p4622,Hatano:2001p4590} but the data have yet to be published.

\subsection{Bolometric Light Curves}\label{bol_lcs}

With an absence of data in the UV and IR, we do not have the photometric coverage to determine true bolometric light curves for SNe 2007bg and 2007bi, but as an approximation we create {\it BVRI} pseudo-bolometric light curves. To determine the {\it BVRI} luminosity for a given set of observations we take the extinction-corrected magnitudes in each band, convert them to flux densities at the effective wavelengths of the standard Johnson-Cousins system and then integrate to get a total flux. To compare the resulting pseudo-bolometric light curves with other SNe, we take {\it BVRI} data from the literature and perform the same conversion on the other SNe. The relative decline rates and absolute {\it R}-band light curve shapes of the various SNe displayed in Figure \ref{RLC-comp-plot} seem to be well duplicated in a comparison of their bolometric light curves displayed in Figure \ref{bolo-comp-plot}. The early decline rate of SN 2007bg is shown to be intermediate to the fast SN 1994I and SN 2002ap and in the tail phase its decline rate is again found to slow to match that of the other SNe Ic-BL. It is to be noted that there may be some remaining flux contamination attributed to the underlying host-galaxy at late times in the {\it B}-band light curve of SN 2007bg that is subsequently integrated into the bolometric-luminosity causing the tail of the light curve to appear shallower than actually it is. 

Also included in the comparison of bolometric light curves is Type IIn SN 2006gy \citep{Smith:2007p158,Ofek:2007p15} which is one of the most luminous SN discovered to date \citep[light curve taken from ][]{Agnoletto:2009p3196}. The luminosity of SN 2006gy is most easily explained by a long-lived radiative shock attributed to the interaction of the ejecta with circumstellar material \citep{Smith:2007p195,Agnoletto:2009p3196,Blinnikov:2008p7366}. Recently \cite{Miller:2009p9538} have found that late-time optical and NIR detections of SN 2006gy are best explained by a scattered light and IR echo resulting from a massive dusty shell which surrounded the progenitor star. Remarkably, after $\sim$70 days SN 2007bi matches the luminosity of SN 2006gy and, with no indication of the narrow emission lines of hydrogen attributed to circumstellar interaction, this extreme luminosity is presumably powered by a huge amount of radioactive $^{56}$Ni being synthesised in the explosion of SN 2007bi.

\begin{figure*}		
\begin{center}
\includegraphics[scale=0.8]{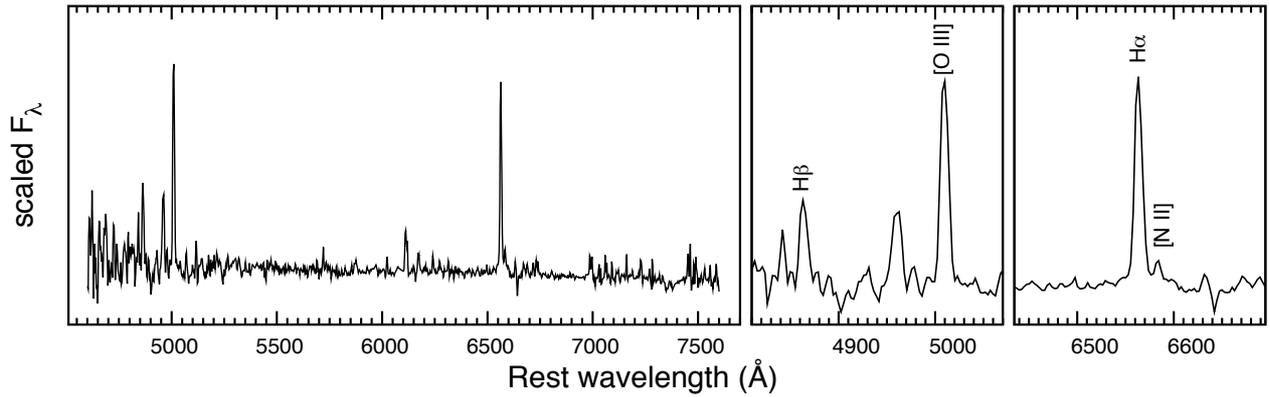}
\caption{Spectrum of the host galaxy of SN 2007bg taken at +767 d relative to the {\it R}-band maximum light of SN 2007bg (2007 April 16). The panels on the right reveal the key emission lines used to determine a host galaxy abundance on the \cite{Pettini:2004p217} scale.}
\label{fig:SN07bi_host_spec}
\end{center}
\end{figure*}

\section{Spectra}\label{spectra}
\subsection{Spectroscopic Observations and Data Reduction}\label{obs_spec} 

Spectroscopic data for SNe 2007bg and 2007bi are reported in Tables \ref{2007bg_spec} and \ref{2007bi_spec} respectively. Our data set includes four spectra of SN 2007bg and four of SN 2007bi, all in the photospheric phase of evolution. Also included is a FORS2 VLT spectrum of SN 2007bi taken at 413 d post {\it R}-band maximum, which we were kindly granted access to by \cite{Gal-Yam:2009}. A spectrum of the host galaxy of SN 2007bg was obtained with the GMOS instrument on the Gemini North Telescope (GNT) more than two years post maximum light in order to obtain an abundance measurement of the galaxy. All frames were reduced (overscan corrected, bias-subtracted and flat-fielded) and the spectra optimally extracted using standard {\tiny IRAF} routines. The extracted spectra were wavelength-calibrated using comparison lamp exposures, and flux-calibrated by using spectroscopic standard-star exposures. Telluric features were identified in reduced standard-star spectra and subsequently removed from the SN spectra. Finally the flux-calibration of the spectra was checked with the SN photometry and corrected for slit-loss or flux lost due to clouds. The spectral evolution of SNe 2007bg and 2007bi are shown in Figures \ref{fig:SN07bg_spectra} and \ref{fig:SN07bi_spectra} respectively, and the spectrum of the host galaxy of SN 2007bg is shown in Figure \ref{fig:SN07bi_host_spec}.

\subsection{Spectral Evolution}\label{spec_evol}

\begin{table*}			
\begin{center}
\caption{\textrm{Spectroscopic observations of SN 2007bg.}\label{2007bg_spec}}
\begin{tabular}{l r r r l l l l l l}
\hline\hline
{UT Date}	& {JD}		& Epoch$^{a}$	& Rest-frame		& Telescope +		& Grisms	& Exp. Time 				& \multicolumn{1}{c}{Range}	& Airmass  \\
		& {- 2,450,000}	& (d)			& epoch (d)$^{a}$	& Instrument$^{b}$ 	& 		&\multicolumn{1}{c}{(secs)}	& \multicolumn{1}{c}{(\AA)}  \\
\hline
21/04/07	& 4211.5 	& 4.8	& 4.6	& TNG + LRS 	& LR-B; LR-R 		& 2 x 2700 	& 3300-9900	& 1.1\\
03/05/07	& 4223.8 	& 17.1	& 16.5	& INT + IDS 	& R150V 			& 2 x 1800 	& 3300-9650 	& 1.1\\
12/05/07	& 4232.6 	& 25.9	& 25.0	& TNG + LRS 	& LR-B; LR-R		& 2 x 2400 	& 3850-8150 	& 1.3\\
15/06/07	& 4266.5	& 59.8	& 57.8	& WHT + ISIS 	& R300B;  R158B		& 900; 1800 & 3200-9500	& 1.6\\
18/06/09	& 5000.3	& 793.6	& 767.0	& GNT + GMOS		& R400				& 2 x 3600	& 4500-7500 & 1.4\\
\hline
\end{tabular}
\end{center}
{\it Notes.} \\
$^{a}$ Relative to the date of discovery, MJD 2\,454\,196.7 (CBET 927) \\
$^{b}$ TNG = 3.58m Telescopio Nazionale Galileo, Observatorio del Roque de Los Muchachos, La Palma (Spain).  INT = 2.5m Isaac Newton Telescope, Observatorio del Roque de Los Muchachos, La Palma (Spain). WHT = 4.2m William Herschel Telescope, Observatorio del Roque de Los Muchachos, La Palma (Spain). GNT = 8.1m Gemini North Telescope, Mauna Kea, Hawaii (USA)\\
\end{table*}
\begin{table*}			
\begin{center}
\caption{\textrm{Spectroscopic observations of SN 2007bi discovered on MJD 2\,454\,197.0}\label{2007bi_spec}}
\begin{tabular}{l r r r l l l l l l}
\hline\hline
{UT Date}	& {JD}		& Epoch$^{a}$	& Rest-frame		& Telescope +		& Grisms	& Exp. Time 				& \multicolumn{1}{c}{Range} & Airmass  \\
		& {- 2,450,000}	& (d)			& epoch (d)$^{a}$	& Instrument$^{b}$ 	& 		&\multicolumn{1}{c}{(secs)}	& \multicolumn{1}{c}{(\AA)}  \\
\hline
22/04/07	& 4213.4 	& 60.4	& 53.6	& Ekar + AFOSC 	& 4; 2		 	& 7200; 3600	& 3100-9000	& 1.3 	\\
24/04/07	& 4215.5	& 62.5	& 55.5	& INT + IDS	 	& R150V		 	& 3600		& 3100-8600	& 1.2	\\
15/06/07	& 4266.5	& 113.5	& 100.7	& WHT + ISIS	 	& R300B; R158B	& 2 x 1800	&  3100-9000	& 1.9\\
21/07/07	& 4303.4	& 150.4	& 133.5	& WHT + ISIS	 	& R158B; R158R	& 2 x 5400	&  3100-9000	& 1.5 - 2.3	\\
10/04/08	& 4566.0	& 413.0	& 366.5	& VLT + FORS2	& GRIS 300V             & 4 x 3600	&  3150-8600	&1.2 - 1.8 \\
\hline
\end{tabular}
\end{center}
{\it Notes.} \\
$^{a}$ Relative to the date of maximum brightness \citep{Gal-Yam:2009}. \\
$^{b}$ Ekar = 1.82m Copernico Telescope, INAF, Osservatorio di Asiago, Mt. Ekar, Asiago (Italy). INT = 2.5m Isaac Newton Telescope, Observatorio del Roque de Los Muchachos, La Palma (Spain). WHT = 4.2m William Herschel Telescope, Observatorio del Roque de Los Muchachos, La Palma (Spain). VLT = 8.2m Very Large Telescope (unit telescope 1), ESO-Paranal Observatory, Cerro Paranal in the Atacama Desert (Chile).\\
\end{table*}

{\it SN 2007bg}: Spectroscopic data for SN 2007bg obtained within the first two months of discovery are shown in Figure \ref{fig:SN07bg_spectra}, ranging from +5 to +58 d (time given in the SN rest-frame). A continuum is present in each of the spectra, revealing that the photospheric phase of evolution lasted for at least two months post {\it R}-band maximum. The first spectrum in the data set (+5 d) displays the broad P Cygni features of Ca {\tiny II} H\&K, Ca {\tiny II} near-IR triplet, Mg {\tiny II} $\lambda$4481, Na {\tiny ID}, Si {\tiny II} $\lambda$6355 and Fe {\tiny II} blends associated with SNe Ic-BL. A weak O {\tiny I} $\lambda$7774 absorption feature also appears to be present. Referring to this spectrum \citet{Harutyunyan:2007p5835} made note of the strong, broad feature centered at about 6524$\AA$ claiming that it may be associated with H$\alpha$. If attributed to H$\alpha$, the position of the minimum at $\sim$6150$\AA$ indicates an expansion velocity of $\sim$18\,000 km s$^{-1}$. Due to the high density of lines in this region of the spectra of SNe Ic  (e.g. Si {\tiny II}, C {\tiny II}, Ne {\tiny I}) the presence of H$\alpha$ is speculative. A stronger argument can be made for the presence the He {\tiny I} situated at a velocity of $\sim$14\,500 km s$^{-1}$. Evidence for the detection of several lines He {\tiny I} lines is present in the optical spectra (see Figure \ref{fig:SN07bg_spectra}). The broad lines of this SN are a consequence of high expansion velocities which are indicative of large energies per unit mass of the SN ejecta. The narrow emission line seen at 5\,400$\AA$ is attributed to the [O {\tiny I}] $\lambda$5577 night sky emission line. In the final spectrum of this sequence (+58 d) there is some indication that forbidden [O {\tiny I}] and possibly semi-forbidden Mg {\tiny I}] (blended with Fe {\tiny II}) emission lines are beginning to become visible. Combined with an obvious increase in the emission component of the Ca {\tiny II} near-IR triplet and the knowledge that at this stage SN 2007bg has clearly settled into the radioactive tail of its light curve (see Figure \ref{bolo-comp-plot}), this would suggest that SN 2007bg is already on the verge of transitioning to the nebular phase. In the GNT spectrum taken + 767 d post {\it R}-band maximum light there is no detection of flux resulting from SN 2007bg.\\

\noindent {\it SN 2007bi}: Spectroscopic data for SN 2007bi obtained in the first three months of discovery are shown in Figure \ref{fig:SN07bi_spectra}, ranging from +54 d to +134 d post {\it R}-band maximum (time given in the SN rest-frame). Also shown is a late time spectrum (+367 d) courtesy of \cite{Gal-Yam:2009}. When compared at similar epochs the spectra of SN 2007bi display expansion velocities lower than that of SN 2007bg, indicating a lower energy per unit mass of ejecta. Individual absorption features such as Fe {\tiny II} $\lambda\lambda$4924, 5018, and 5169 (multiplet 42) that are blended in the spectra of SN 2007bg are individually resolved in the early spectra of SN 2007bi. Again P Cygni features of Ca {\tiny II} H\&K, Ca {\tiny II} near-IR triplet, Na {\tiny ID},  and Si {\tiny II} $\lambda$6355 are present. There are multiple, narrow Fe {\tiny II} absorption features within the wavelength range 3900-5500$\AA$; the most prominent of which is the aforementioned blueshifted Fe {\tiny II} multiplet 42. There is probably also a contribution the from Mg {\tiny II} $\lambda$4481 absorption line within this region and tentatively also several narrow Ti {\tiny II} absorption lines. The absorption feature located within the wavelength range 7300-7900$\AA$ is probably due to a blend of blueshifted O {\tiny I} $\lambda$7774 and Mg {\tiny II} $\lambda$7890. The most prominent feature displayed in the spectra of SN 2007bi also happens to be the most peculiar, that is the emission feature centered around 7320$\AA$. This feature is presumably associated with the emission of the forbidden [Ca {\tiny II}] $\lambda\lambda$7291, 7324 doublet. How this nebular feature is so pronounced within the clearly photospheric phase of evolution remains to be determined (see Section \ref{spec_comp}). Another surprising characteristic of SN 2007bi is the extreme longevity of its spectral evolution; at +367 d post-discovery the SN still retains a continuum and is only beginning its transition to the nebular phase with the appearance of the forbidden [O {\tiny I}] and semi-forbidden Mg {\tiny I}] features.

\subsection{Comparison of Spectra with those of other SNe}\label{spec_comp}

\begin{figure}[t]		
\begin{center}
\includegraphics[scale=0.73, angle=270]{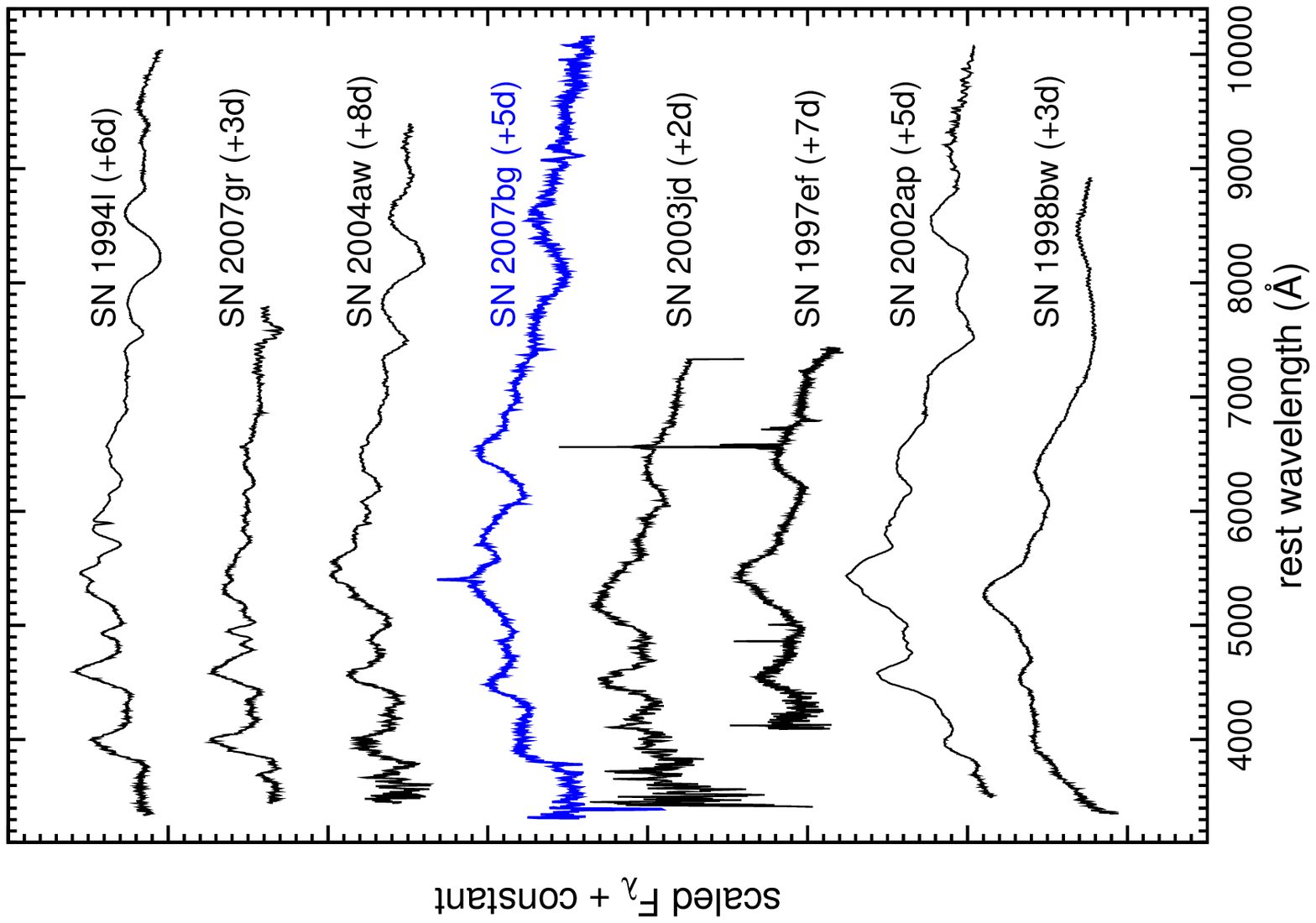}
\caption{SN Ic spectrum comparisons at $\sim$5 days post {\it R}-band maximum. All comparison data have been corrected for extinction. Epochs are relative to the {\it R}-band maximum light and in the SN rest-frame. {\it References and color-excess:} SN 1994I with $E(B-V)=0.300$ \citep{Millard:1999p4678}, SN 2007gr with $E(B-V)=0.092$ \citep{Valenti:2008p233,Hunter:2009p13049}, SN 1997ef with $E(B-V)=0.042$ \citep{Iwamoto:1998p4660,Mazzali:2000p4496}, SN 1998bw with $E(B-V)=0.065$ \citep{Patat:2001p4674}, SN 2002ap with $E(B-V)=0.09$ \citep{GalYam:2002p5310, Foley:2003p136}, SN 2003jd with $E(B-V)=0.14$ \citep{Valenti:2008p66, Mazzali:2005p5329} and SN 2004aw with $E(B-V)=0.37$ \citep{Taubenberger:2006p4726}. All spectra are shown in the rest-frame of the SN and are placed in order of increasing photospheric velocity from top to bottom.}
\label{fig:5d_spectra_comp_fig}
\end{center}
\end{figure}
\begin{figure}[t]		
\begin{center}
\includegraphics[scale=0.73, angle=270]{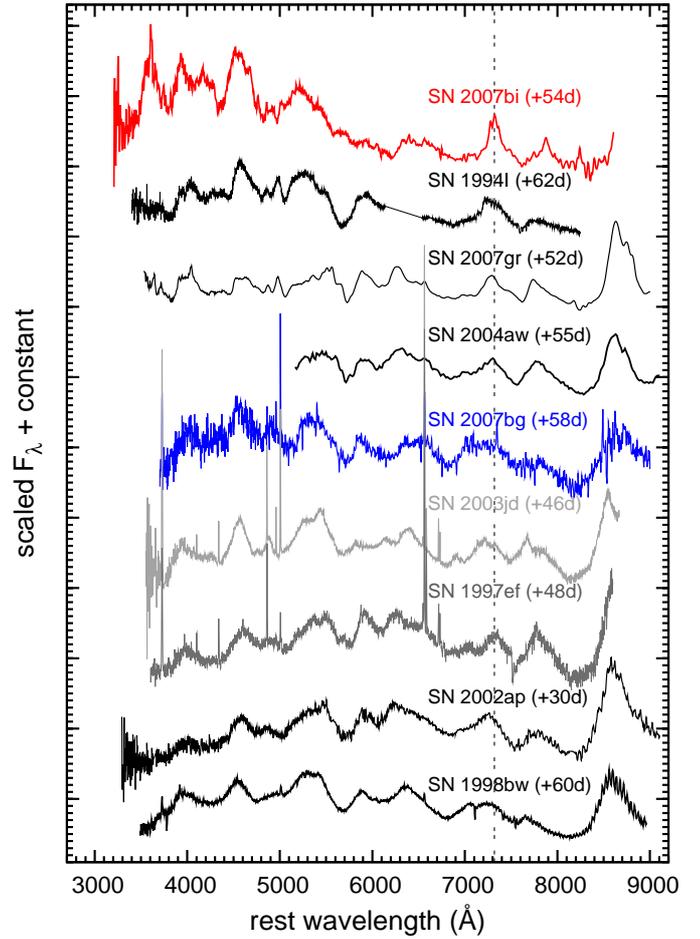}
\caption{SN Ic spectrum comparisons at $\sim50$ days post {\it R}-band maximum. All comparison data have been corrected for extinction. Epochs are relative to the {\it R}-band maximum light and in the SN rest-frame. References as in Figure \ref{fig:5d_spectra_comp_fig}. The dashed line marks the position of the [Ca {\tiny II}] feature prominent in the spectra of SN 2007bi.}
\label{fig:60d_spectra_comp_fig}
\end{center}
\end{figure}
\begin{figure}[h]		
\begin{center}
\centering
\includegraphics[scale=0.73, angle=270]{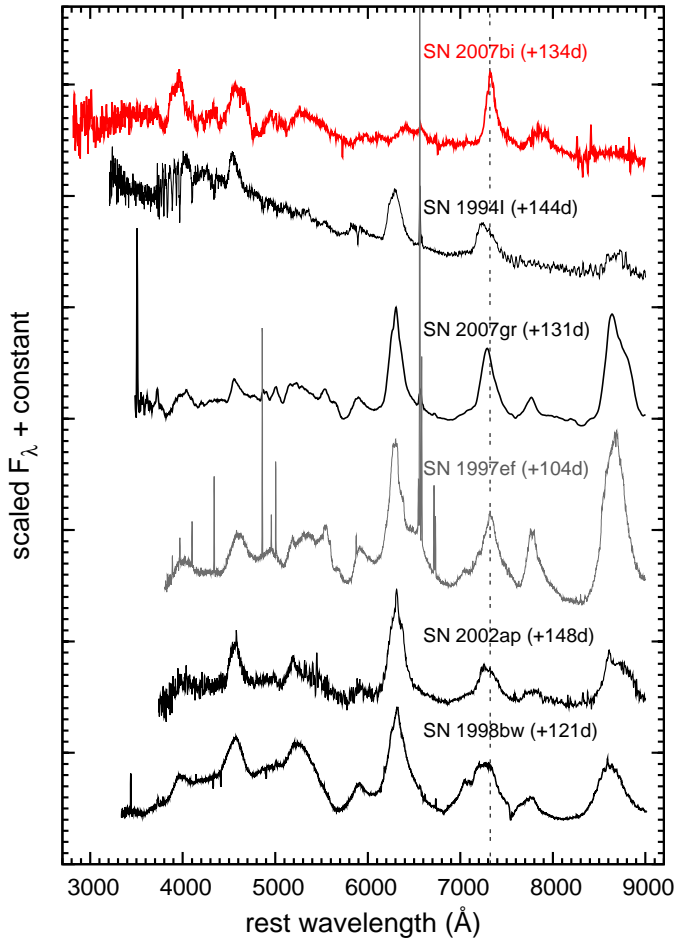}
\caption{SN Ic spectrum comparisons at $\sim4$ months post {\it R}-band maximum. All comparison data have been corrected for extinction. Epochs are relative to the {\it R}-band maximum light and in the SN rest-frame. References as in Figure \ref{fig:5d_spectra_comp_fig}. The dashed line marks the position of the [Ca {\tiny II}] feature prominent in the spectra of SN 2007bi. Note that the blue-continuum seen in the spectrum of SN 1994I is due to background light contamination.}
\label{fig:120d_spectra_comp_fig}
\end{center}
\end{figure}
\begin{figure}[h]		
\begin{center}
\includegraphics[scale=0.73, angle=270]{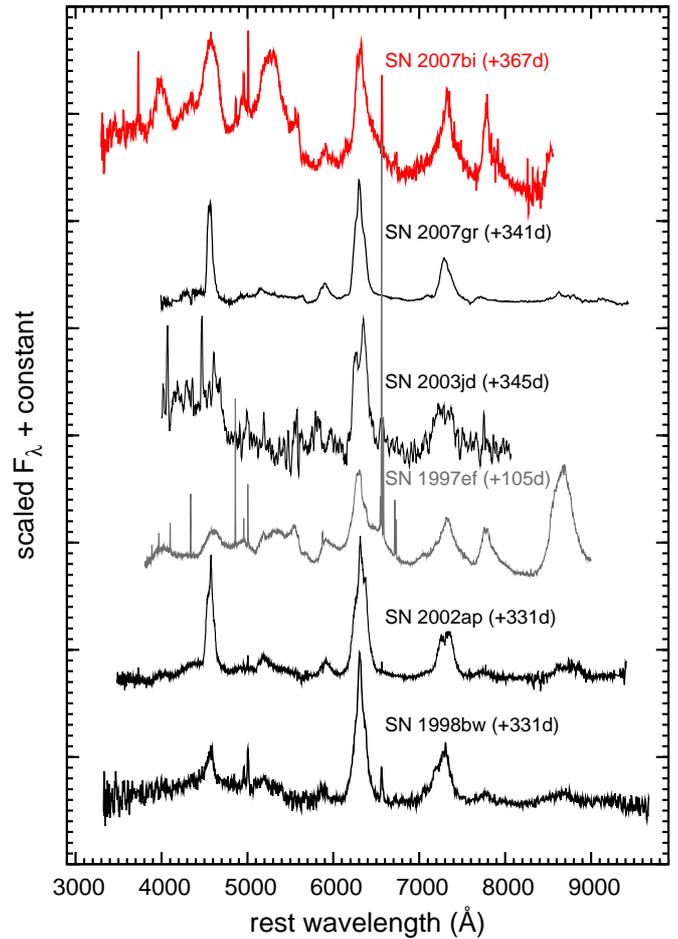}
\caption{SN Ic spectrum comparisons at approximately one year post {\it R}-band maximum. All comparison data have been corrected for extinction. Epochs are relative to the {\it R}-band maximum light and in the SN rest-frame. References as in Figure \ref{fig:5d_spectra_comp_fig}.}
\label{fig:330d_spectra_comp_fig}
\end{center}
\end{figure}
\begin{figure}[h]		
\begin{center}
\includegraphics[scale=0.73,angle=270]{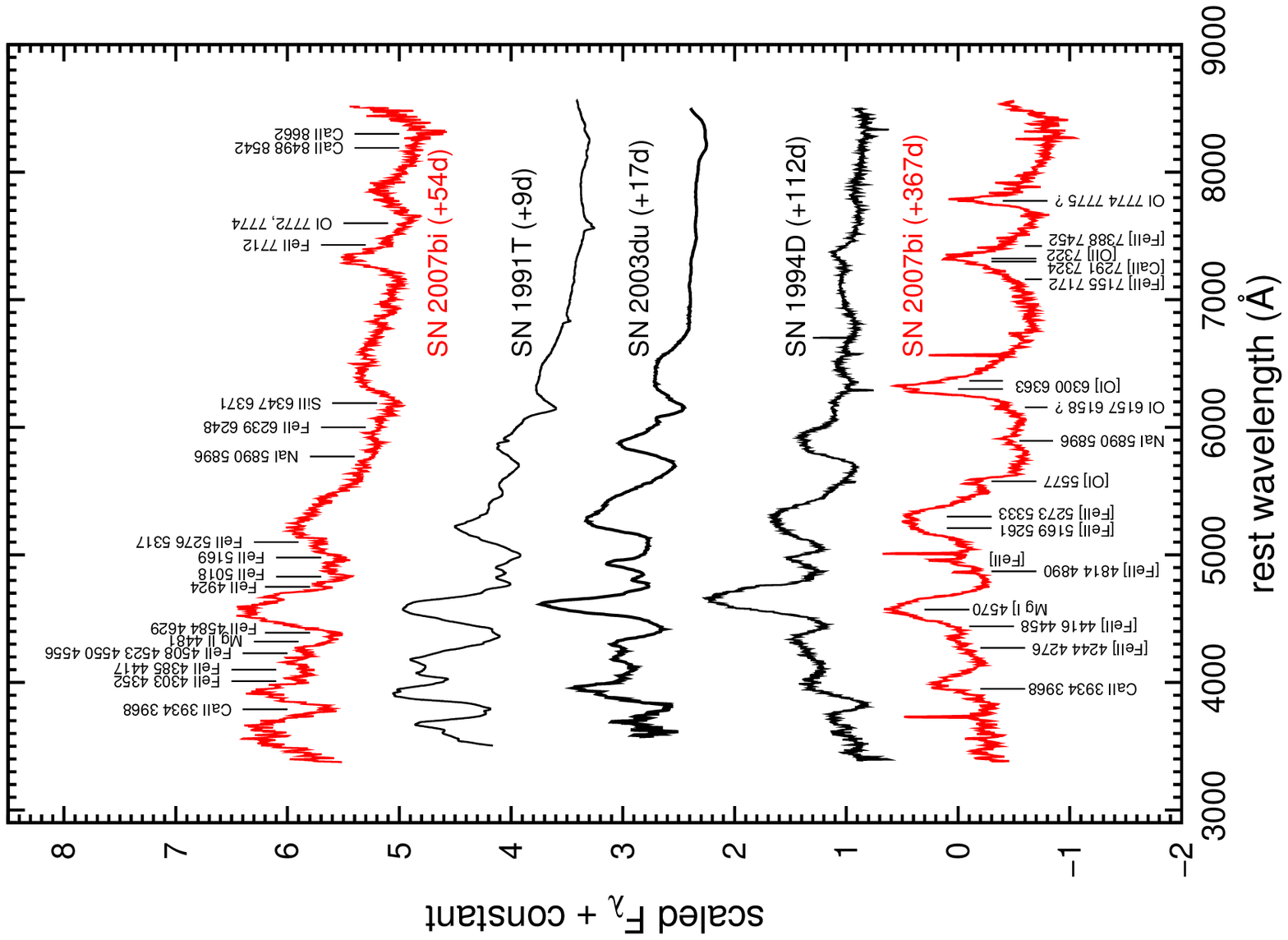}
\caption{SN 2007bi compared with SNe Type Ia. The prominent iron features present in the spectra of SN 2007bi, similar to those seen in SNe Type Ia, are indicative of a large $^{56}$Ni mass synthesised in the ejecta of SN 2007bi. All SNe shown are corrected for extinction and displayed within their respective rest-frames. SNe 1991T with $E(B-V)=0.14$ \citep{Phillips:1999p10447} \citep{Mazzali:1995p11574}, 2003du with $E(B-V)=0.01$ \citep{Anupama:2005p7822,Stanishev:2007p7824} and 1994D with $E(B-V)=0.042$ \citep{Meikle:1996p7832,Turatto:1996p7865}.}
\label{fig:Ia_comp_fig}
\end{center}
\end{figure}

The spectra of SNe 2007bg and 2007bi are compared with other SNe Ic at similar epochs in Figures \ref{fig:5d_spectra_comp_fig} to \ref{fig:330d_spectra_comp_fig}. For the purpose of clarity the spectra are arranged from top to bottom in order of increasing line velocity (as stated in the literature). The Generic Classification Tool (GELATO\footnote{https://gelato.tng.iac.es}) developed by \cite{Harutyunyan:2008p8386}, finds that the first spectrum of SN 2007bg (+5 d) is most readily associated with that of SN 1998bw. This is mainly due to the broad absorption features common to both objects at early-times. These broad features, also seen in SNe Ic-BL 2002ap, 1997ef, 2003jd and to a lesser extent SN 2004aw, suggest that a significant amount of the matter from the SN explosion was ejected at unusually high velocities; greater than $\sim$14\,000 km s$^{-1}$ in the case of SN 2007bg (see Section \ref{phot_vel}).

At  +5 d SN 2007bg joins SN 2002ap and SN 1998bw in displaying a clear lack of flux blueward of $\sim$4\,500$\AA$ due to strong line blanketing in this region (see Figure \ref{fig:5d_spectra_comp_fig}). This is certainly not the case for SN 2007bi which reveals an abundance of iron group features in this spectral region and continues to do so throughout the course of its photospheric evolution (see Figure \ref{fig:60d_spectra_comp_fig}). In the spectra of SNe Ic the absorption feature associated with Na {\tiny ID} generally becomes increasingly more distinct with time. However this feature is already very distinct in the early-time spectra of SN 2007bg which suggests a contamination by other spectral ions, the most likely of which is He {\tiny I} $\lambda$5876.

The dashed line in Figures \ref{fig:60d_spectra_comp_fig} and \ref{fig:120d_spectra_comp_fig} marks the location of the forbidden [Ca {\tiny II}] feature usually not seen until a SN begins to transition from the photospheric to the nebular phase of evolution when the density of the ejecta is low enough to allow for this feature. Although it is surprising that this nebular feature appears this early in the evolution of SN 2007bi, it is not unprecedented in the spectra of other SNe Ic. As can be seen in Figure \ref{fig:60d_spectra_comp_fig}, at $\sim$50 d [Ca {\tiny II}] is clearly beginning to emerge in the spectra of SNe 2004aw, 1994I and 1997ef. By $\sim$130 d the feature has become pronounced in the spectra of all SNe compared (see Figure \ref{fig:120d_spectra_comp_fig}). Considering the extreme longevity of the spectral evolution of SN 2007bi it is still unclear as to how this forbidden feature can appear so distinguished at such an early stage in the lifetime of this SN.

The spectral evolution of SN 2007bi is astonishingly slow when compared to other SNe Ic, mimicking the very slow post-maximum decline of its light curves. When SNe 1998bw, 2002ap, 2003jd and 2007gr have all settled into the nebular phase of evolution displaying forbidden emission features of [O {\tiny I}], [O {\tiny II}], [Ca {\tiny II}] and semi-forbidden Mg {\tiny I}], at +367 d SN 2007bi still reveals a continuum and strong photospheric lines with nebular features super-imposed on top (see Figure \ref{fig:330d_spectra_comp_fig}).

\begin{figure*}		
\begin{center}
\includegraphics[scale=0.6, angle=270]{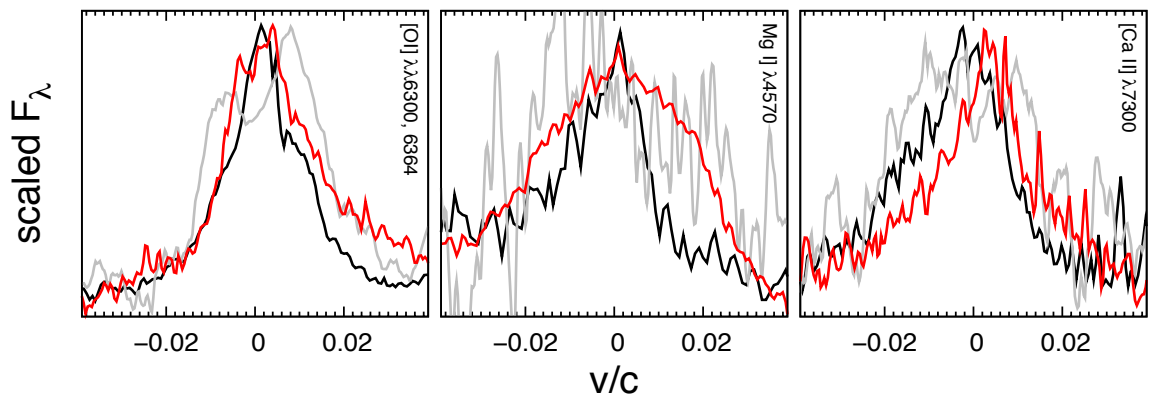}
\caption{A comparison of the profiles of the [O {\tiny I}], Mg {\tiny I}] and [Ca {\tiny II}] emission lines found in the nebular spectra of SNe 1998bw (black), 2003jd (grey) and the final + 367 d spectrum of SN 2007bi (red).
\label{fig:fOI_fig}}
\end{center}
\end{figure*}

Figure \ref{fig:fOI_fig} shows the profiles of the [O {\tiny I}], Mg {\tiny I}] and [Ca {\tiny II}] observed in latest spectrum of SN 2007bi (red) compared to that of SNe 1998bw (black) and 2003jd (grey) when observed in the nebular phase. The double peak in the [O {\tiny I}] profile of SN 2003jd has been used as evidence for an asymmetric explosion \citep[as with many other SN Ibc ][]{Maeda:2008p7636,Modjaz:2008p2993,Taubenberger:2009p9820,Milisavljevic:2009p8419} which \cite{Mazzali:2005p5329} claim as evidence for an off-axis GRB ($\gtrsim70^{o}$ from our line-of-sight)\footnote{However in the specific case of SN 2003jd, an off-axis GRB can probably be ruled out as the object did not exhibit the radio emission expected at late-time from a GRB afterglow \citep{Soderberg:2005p9584}.}. The [O {\tiny I}] profile of SN 2007bi is not double-peaked, revealing no evidence of highly-asymmetric explosion. 

Due to the large amounts of iron in the spectra of SN 2007bi, it is beneficial to compare this object with the spectra of a few Type Ia SNe. Figure \ref{fig:Ia_comp_fig} displays SN 2007bi compared with three such SNe; SNe 1991T \citep{Mazzali:1995p11574}, 2003du \citep{Anupama:2005p7822} and 1994D \citep{Turatto:1996p7865, Meikle:1996p7832}. Blueward of $\lambda5500$ the + 367 d spectrum of SN 2007bi reveals an extraordinary similarity to the + 112 d spectrum of SN 1994D with a multitude of forbidden Fe {\tiny II} lines. As the daughter element of $^{56}$Co, itself the daughter of $^{56}$Ni, this abundance of $^{56}$Fe is the unmistakable by-product of a huge amount of $^{56}$Ni synthesised in the explosion of SN 2007bi. It is the dominance of oxygen features redwards of 6\,000$\AA$ that clearly distinguishes this event from SNe Type Ia. For the + 54 d spectrum of SN 2007bi we overplot the line identifications for the Type Ia SN 1990N \citep{Mazzali:1993p7880} and for the + 367 d spectrum we overplot the line identifications from SN 1998bw \citep{Mazzali:2001p6332}. The agreement of most of the line identifications confirms the large amount of iron present in the spectra.

The strength of both the oxygen and iron lines at the same time are unprecedented in any Type I SN, pointing towards a massive CO core that exploded producing a large mass of $^{56}$Ni.

\section{Other SNe Parameters}\label{other_para}

\subsection{Photospheric Velocities}\label{phot_vel}

\begin{figure}[t]		
\begin{center}
\centering
\includegraphics[scale=0.45, angle=270]{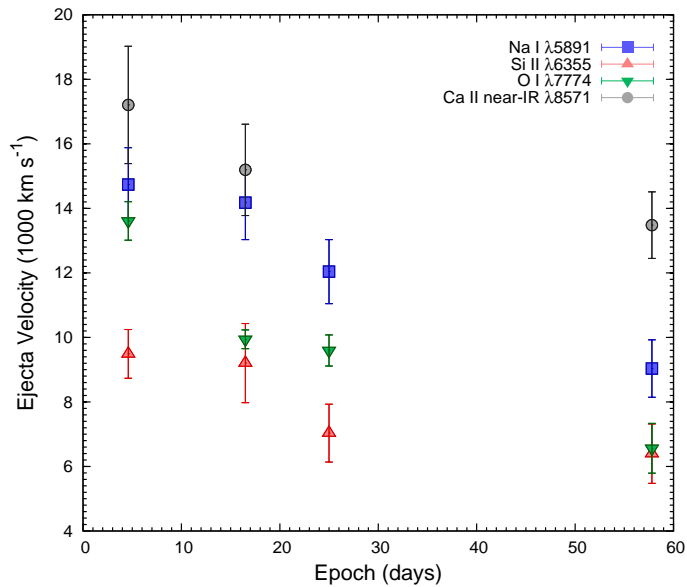}
\caption{Ejecta velocities of SN 2007bg. Time given in the SN rest-frame.}
\label{fig:2007bg_evs}
\end{center}
\end{figure}

\begin{figure}[t]		
\begin{center}
\includegraphics[scale=0.45, angle=270]{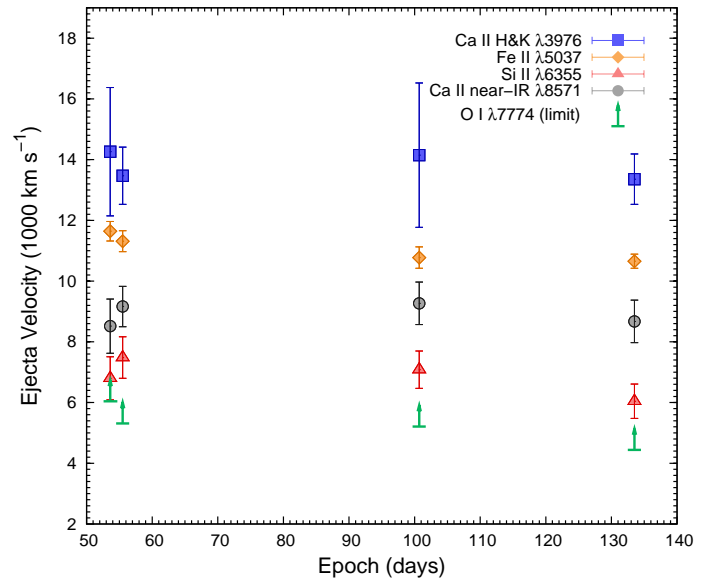}
\caption{Ejecta velocities of SN 2007bi. Time given in the SN rest-frame. Note that the velocity of O {\tiny I} $\lambda7774$ measured in the spectra of SN 2007bi can only be treated as a lower-limit; a consequence of contamination in the blue-wing by the peculiar [Ca {\tiny II}] feature.}
\label{fig:2007bi_evs}
\end{center}
\end{figure}

We measure the velocities of certain spectral ions found in the ejecta of SNe 2007bg and 2007bi by fitting Gaussian profiles to the absorption components of the P Cygni features in rest-frame spectra. The blue-shifted minima of these Gaussian profiles are used to estimate the expansion velocities of the ejected material. As a caveat, many of these features may actually be a blend of multiple lines, especially with the higher ejecta velocities seen in SN 2007bg. Figure \ref{fig:2007bg_evs} shows the velocity evolutions of Na {\tiny ID}, O {\tiny I} $\lambda$7774, Si {\tiny II} $\lambda$6355 and the Ca {\tiny II} near-IR triplet seen in SN 2007bg. Here we find that the Ca {\tiny II} near-IR triplet has consistently the highest velocity and Si {\tiny II} the lowest velocity, with O {\tiny I} and Na {\tiny ID} intermediate (note that Na {\tiny ID} feature may be contaminated by He {\tiny I} $\lambda$5876).  Figure \ref{fig:2007bi_evs} shows the velocity evolution of Ca {\tiny II} H$\&$K, Fe {\tiny II} triplet ($\lambda\lambda$4924, 5018, and 5169), Si {\tiny II} $\lambda$6355 and O {\tiny I} $\lambda$7774 in SN 2007bi. Expansion velocities of $\sim$12\,000 km $s^{-1}$ of the  Fe {\tiny II} triplet measured at + 47 and + 48 days post {\it R}-band maximum \citep{Nugent:2007p3645}, combined with our measurements reveal the consistency of the photospheric velocities of this object over at least a $\sim90$ day period. Given that this object was discovered more than 40 days post {\it R}-band maximum light, this consistency is not so peculiar, as generally the photospheric velocities drop steeply in the first few days post-maximum before plateauing off to a steady velocity. However to find Ca {\tiny II} and Fe {\tiny II} at such high velocities at these late times is unusual. Note that the velocity of O {\tiny I} $\lambda7774$ measured in the spectra of SN 2007bi can only be treated as a lower-limit, as a consequence of contamination in the blue-wing by the [Ca {\tiny II}] feature.

\subsection{A Search for associated GRBs}\label{GRB?}

To identify whether or not there was an associated GRB detected for either SNe 2007bg or 2007bi we searched the GRBlog\footnote{http://grblog.org} database \citep{Quimby:2004p7563} to determine if any of the seven high-energy satellites in operation at the time detected an event in the direction of either SNe. The seven satellites in operation at the time were the {\it Mars Observer}, {\it Wind-Konus}, RHESSI, INTEGRAL, {\it Swift}, MESSENGER and {\it Suzaku}.  A total of 5 bursts were observed over the period {\tiny UT} 2007 March 27 to April 11 (5 - 20 d prior to discovery of SN 2007bg). All bursts were localised and not coincident with SN 2007bg within the respective errors of their positions. \cite{Soderberg:2007p5839} also observed the field of SN 2007bg with the Swift X-Ray Telescope (XRT) on 2007 Apr 22.4 ($\sim$6 d post {\it R}-band maximum light), but did not detect an X-ray source coincident with the SN to a 3$\sigma$ upper-limit of $F_{x} = 4.6\times10^{-14}$ ergs s$^{-1}$ cm$^{-2}$  ($0.2-10$ keV) at the distance of the SN. 

A total of 11 confirmed bursts and 1 possible burst (possibly a cosmic ray shower) were observed over the period {\tiny UT} 2007 Jan 1 to Feb 15 (6 - 51 d prior to R-band maximum light of SN 2007bi). Again all bursts were localised but none were found coincident with SN 2007bi within the respective errors of their positions. As no burst was found to be coincident with either SNe 2007bg or 2007bi we assume that neither SN had an associated GRB collimated {\it along our line-of-sight}. This leads to the following possible conclusions: ({\it a}) neither SN was associated with a GRB, ({\it b}) the GRB was beamed at an angle away from our line-of-sight or ({\it c}) the GRB was simply too weak to have been detected by any of the aforementioned instruments (or a combination of {\it b} and {\it c}).

Regardless of the initial projection angle of a GRB jet, it will be decelerated to sub-relativistic speeds as it sweeps up and shocks surrounding circumstellar material (CSM). Once this happens the jet fails to remain collimated and begins to spread sideways until eventually the GRB afterglow emission is visible from all viewing angles. The specific point in time a GRB afterglow becomes visible occurs when the decelerated jet finally spreads into our line-of-sight and depends on the initial projection angle, the energy of the jet and the density profile of the surrounding CSM. This may be weeks to years post-explosion and is generally announced by a sharp rise in synchrotron emission at the location of the explosion, typically best observed in the radio band. A detection of late-time radio emission at the location of a SN Ibc (including SNe Ic-BL) would therefore provide evidence (but not proof) for an off-axis GRB.

As the host galaxy of SN 2007bg is of remarkably low luminosity and presumably metal-poor (see Section \ref{hosts}) \citet{Prieto:2007p5838} flagged SN 2007bg as a prime candidate for having an off-axis GRB. It has recently been revealed that within the first two years of its evolution SN 2007bg has shone remarkably bright in the radio-band (\citealt{Prieto:2009p6880}, GCN 9444; \citealt{Soderberg:2009p9094}, ATel 2066) with luminosities surpassing those of Type Ic SN 2003bg \citep{Soderberg:2006p7066,Hamuy:2009p12161,Mazzali:2009p12162} and Type Ic SN 2003L \citep{Soderberg:2005p9584}. Analysing three epochs of publicly available data from the Very Large Array (VLA), \citet{Prieto:2009p6880} state that SN 2007bg has radio luminosities $\sim$3, 125 and 315 days post-discovery of $3.6\times10^{27}$, $1.4\times10^{28}$ and $4.4\times10^{28}$ ergs s$^{-1}$ Hz$^{-1}$ at 8.46 GHz. \cite{Prieto:2009p6880} state that these observations of strong radio-emission from early through to late-times in the evolution of SN 2007bg strengthen the case of an off-axis GRB, especially considering that all other SNe Ic-BL observed in the lowest-metallicity environments have been associated with a GRB \citep{Modjaz:2008p87}.

However, radio emission from SNe Ibc can also be generated through the dynamic interaction of fastest moving (mildly relativistic) ejecta with CSM, without the need of a GRB. \citet{Soderberg:2009p9094} explain the radio emission resulting from SN 2007bg in this fashion, concluding that the presence of a GRB does not necessarily have to be invoked to explain the detection of radio-emission from SN 2007bg.

\subsection{Explosion Parameters}\label{ejected_mass}

Modeling of the bolometric light curve of a SN can be used to estimate various explosion parameters, namely the total ejected mass $M_{ej}$, the ejected $^{56}$Ni mass $M_{Ni}$ and the total kinetic energy $E_K$. Making use of the two-component toy model for SNe Ic-BL, first conceived by \cite{Maeda:2003p8320} and later implemented by \cite{Valenti:2008p66} to model the light curve of SN 2003jd, we are able to estimate these parameters for SN 2007bg. The model assumes that the ejecta of SNe Ic-BL can be defined as having two components: a high-velocity, low-density outer-component which influences the shape of the light curve near peak, and a dense inner-component that informs the linear decline of the light curve after peak. As the {\it BVRI} pseudo-bolometric light curve of SN 2007bg is similar to that of SN 2002ap, we approximate the missing UV and IR components of the SN 2007bg light curve by assuming the fractional luminosities contributed by these components is same as measured for SN 2002ap. Using the photospheric velocity measured near maximum brightness to break the degeneracy in the estimation of $M_{ej}$ and $E_K$,  we derive  $E_K=4{\tiny\pm1} \times10^{51}$ergs, $M_{ej}=1.5\pm0.5{\rm M}_\odot$ and $M_{Ni}=0.12\pm0.02{\rm M}_\odot$ for SN 2007bg. The fast decline of the light curve and fast spectroscopic evolution of this object can be seen to be result of its high kinetic energy to ejected mass ratio ($E_K/M_{ej}\sim 2.7$). 

\begin{table}[t]			
\begin{center}
\caption{\textrm{Emission lines used to measure abundances of the host galaxies of SNe 2007bg and 2007bi.}\label{abundances}}
\begin{tabular}{l l l}
\hline\hline
{Emission Line}	& {SN 2007bg} & {SN 2007bi} \\
\hline
$[$O {\scriptsize II}$]$ $\lambda3727$ 	&  $-$						& 3.75 {\tiny $\pm$ 0.93} \\
H$\beta$ $\lambda4861$ 					& 6.19 {\tiny $\pm$ 0.69} 	& 1.20 {\tiny $\pm$ 0.45} \\
$[$O {\scriptsize III}$]$ $\lambda4959$ & 8.45 {\tiny $\pm$ 0.71} 	& 1.75 {\tiny $\pm$ 0.95} \\
$[$O {\scriptsize III}$]$ $\lambda5007$ & 17.81 {\tiny $\pm$ 1.82} 	& 3.53 {\tiny $\pm$ 1.07} \\
H$\alpha$ $\lambda6563$ 				& 16.01 {\tiny $\pm$ 1.62} 	& 3.33 {\tiny $\pm$ 0.84} \\
$[$N {\scriptsize II}$]$ $\lambda6584$ 	& 0.91 {\tiny $\pm$ 0.13} 	& $<0.43$ \\
\hline
\end{tabular}
\end{center}
{\it Notes.} \\
Emission line fluxes in units 10$^{-17}$ erg s$^{-1}$ cm$^2$. Errors quoted are the uncertainties measured when fitting Gaussian profiles to the emission lines of the host galaxies. For the host of SN 2007bi an additional component is included in the errors to account for the uncertainty of subtracting off the SN continuum (set at a conservative 20\% of the measured line flux). \\
\end{table}

Unfortunately without photometric coverage around the light curve peak of SN 2007bi we are unable to give an accurate estimation of either $M_{ej}$ or $E_K$, but the slow decline of the light curve and the slow spectroscopic evolution leads us to speculate that the $E_K/M_{ej}$ ratio is relatively low (opposite to SN 2007bg). From the peak magnitude measured by \cite{Gal-Yam:2009} and the tail of the light curve we are able to roughly estimate $M_{Ni}=3.5-4.5 {\rm M}_\odot$ (assuming that all of the luminosity generated by the radioactive decay of $^{56}$Ni).

\subsection{Host galaxies}\label{hosts}

\begin{figure}[t]		
\begin{center}
\includegraphics[scale=0.43]{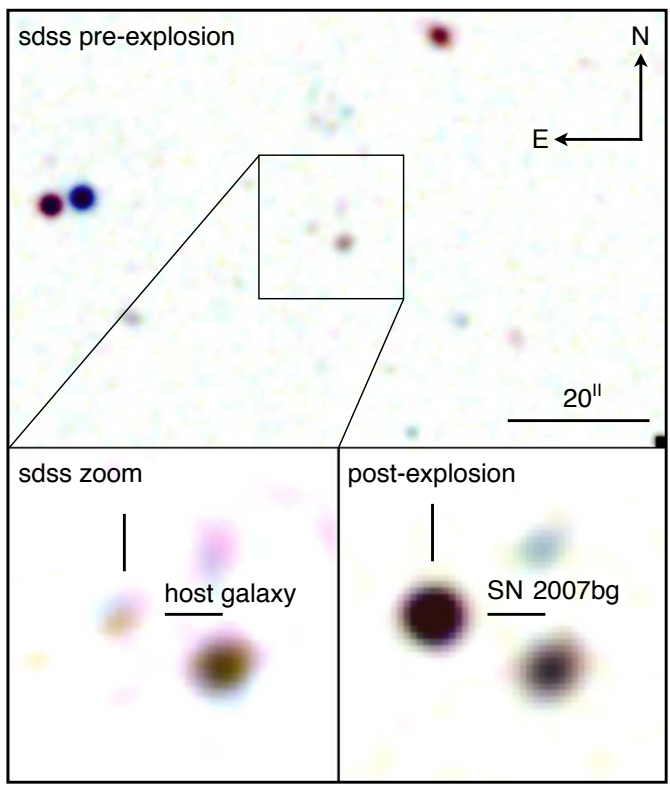}
\caption{SN 2007bg and its host galaxy. The {\it top segment} is a {\it gri}-colour image with frames taken from the Sloan Digital Sky Survey. {\it The bottom-left} segment is an expanded view revealing the extremely sub-luminous host of SN 2007bg. The galaxy to the south-west of the anonymous host is SDSS J114925.74+514920.2 \citep{Prieto:2007p5838}. The {\it bottom-right} segment is a {\it BVR}-colour image of SN 2007bg created using the images taken on the 2007 May 11th by the 2.0m Liverpool Telescope. Image colour has been inverted for clarity.}
\label{fig:2007bg_host_fig}
\end{center}
\end{figure}

SN 2007bg is hosted by an extremely sub-luminous, anonymous galaxy (see Figure \ref{fig:2007bg_host_fig}). This anonymous host is located 1.6$\arcsec$ north and 4.0$\arcsec$ east of {\it SDSS J114925.74+514920.2}, a source morphologically identified as a galaxy in the SDSS-DR7 catalog \citep{Prieto:2007p5838, Prieto:2008p239}. The photometry of the host galaxy has not been measured within any of the SDSS data releases. To determine the photometry we take the SDSS {\it gri}-band images of the field including the host galaxy, together with the SDSS derived {\it gri} photometry of the SN 2007bg local standard stars (Table \ref{2007bg_sequence}) from the SDSS DR7 archive. We perform PSF-fitting photometry\footnote{We consider the host-galaxy to be spatially compact enough to be considered a point-source.} on the galaxy to calculate {\it gri} magnitudes and then use the transformation of \cite{Smith:2002p2474} to determine that the host galaxy has a magnitude $B=23.5${\tiny$\pm0.6$}. The distance to SN 2007bg and Galactic extinction in the line-of-sight (Section \ref{redshifts}) leads to an absolute magnitude of $M_{B}=-12.4$ mag. This is remarkably sub-luminous, which implies that this galaxy is of extremely low-metallicity. By comparison the proto-typical low-luminosity, blue, irregular galaxy I Zw 18 has an absolute magnitude of $M_{B}=-14.45$ mag and an oxygen abundance of only 12+log(O/H) = 7.24 \citep{GildePaz:2003p8017,McGaugh:1991p7985}.

\begin{figure}[t]
\begin{center}
\includegraphics[scale=0.43]{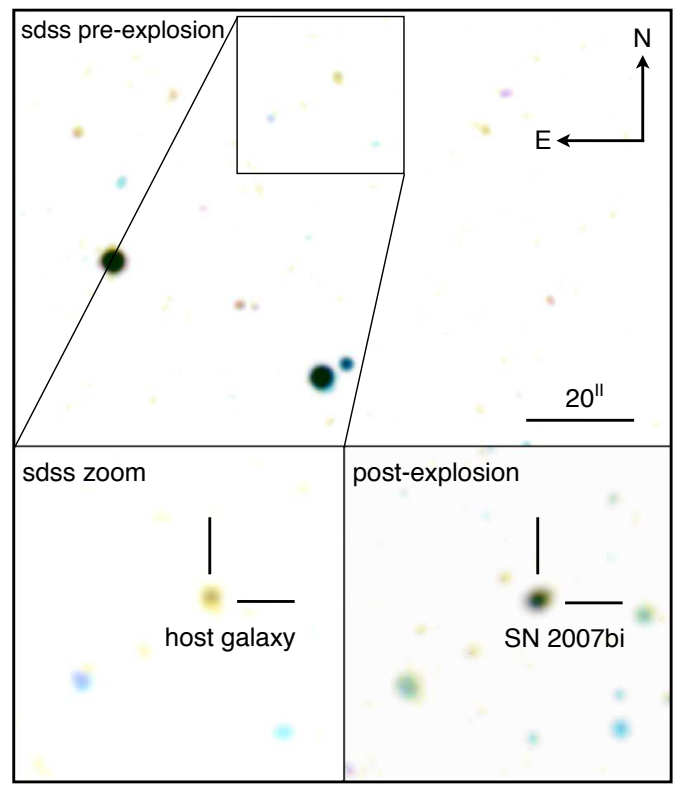}
\caption{SN 2007bi and its host galaxy. The {\it top segment} is a {\it gri}-colour image with frames taken from the Sloan Digital Sky Survey. The {\it bottom-left} segment is an expanded view revealing the sub-luminous host of SN 2007bi. The {\it bottom-right} segment is a {\it BVR}-colour image of SN 2007bi created using the images taken on the 2008 April 10th by the FORS2 instrument on the Very Large Telescope 8.2m unit Telescope 1. Image colour has been inverted for clarity.}
\label{fig:2007bi_host_fig}
\end{center}
\end{figure}

SN 2007bi is coincident with the SDSS identified galaxy {\it SDSS J131920.14+085543.7} (see Figure \ref{fig:2007bi_host_fig}) which has magnitude $B=22.5${\tiny$\pm0.2$}; converted from the SDSS magnitude system with the \cite{Smith:2002p2474} transformation. Using the distance to, and Galactic extinction in the line-of-sight of SN 2007bi leads to an absolute magnitude of $M_{B}=-16.4$ mag. Although not as sub-luminous as either the host of SN 2007bg or I Zw 18, the host of SN 2007bi still exhibits a relatively low luminosity and presumably has a moderately low metallicity. The uncertainty in the photometry of the host galaxies of both SNe 2007bg and 2007bi prohibit their spectral energy distributions from being constrained and K-corrections from being calculated \citep{Blanton:2003p13044}. We caution that these K-corrections may be significant, especially at the moderate redshift of SN 2007bi.

\cite{Modjaz:2008p87} have compared the oxygen abundances at the sites of a sample of 12 nearby non-GRB/SNe Ic-BL with abundances measured at the sites of 5 nearby GRB/SNe Ic-BL. Displaying their results within three different abundance scales \citep{McGaugh:1991p7985,Kewley:2002p8029,Pettini:2004p217} they found that the GRB/SNe Ic-BL were consistently found to be hosted in lower-metallicity environments than {\it all} non-GRB/SNe Ic-BL.

As SN 2007bg is a SN Ic-BL that was not observed to be associated with a GRB (see Section \ref{GRB?}) it is interesting to determine exactly where its host galaxy lies relative to the sample of \cite{Modjaz:2008p87}. Although SN 2007bi is not classified as a SN Ic-BL and probably resulted from a different explosion mechanism to the SNe Ic-BL considered here (see Section \ref{pisn}), it was a hugely energetic explosion \citep[\mbox{$E_{K}\approx10^{53}$ergs},][]{Gal-Yam:2009} discovered in a sub-luminous dwarf galaxy analogous to those that host GRB/SNe Ic-BL.

To this end it is also of value to determine where SN 2007bi lies with respect to the \cite{Modjaz:2008p87} sample. \cite{Pettini:2004p217} (PP04) make use of the [O {\tiny III}] $\lambda 5007$ to [N {\tiny II}] $\lambda 6583$ nebular emission-line ratio (O3N2) to empirically place oxygen abundances on the direct electron temperature ($T_e$) scale. Measuring the H$\beta$, [O {\tiny III}], H$\alpha$ and [N {\tiny II}] line strengths from the GNT spectrum of the host galaxy of SN 2007bg, we calculate the galaxy abundance to be 12+log(O/H$)_{\rm PP04}=8.18${\tiny $\pm0.17$}. This moderately low-metallicity seems at odds with the extremely low metallicity expected from the remarkably low luminosity of the galaxy. In the latest spectrum of SN 2007bi, the narrow [O {\tiny III}] emission line from the underlying host galaxy is clearly distinguishable, but the [N {\tiny II}] line is not. We therefore cannot measure the oxygen abundance of the host of SN 2007bi using the PP04 calibration directly. The position of SN 2007bg on the PP04 / $T_e$ abundance-luminosity plot is shown in Figure \ref{fig:PP04}. The red squares represent GRB/SNe Ic-BL and the blue circles represent non-GRB/SNe Ic-BL, now including SN 2007bg. The pink line indicates the magnitude of the host of SN 2007bi. \cite{Modjaz:2008p87} use the PP04 calibration to place all of the non-GRB/SNe Ic-BL on this plot, but use the $T_e$ method to derive the abundances of the GRB/SNe Ic-BL. 

It has been suggested that at low-metallicities the PP04 calibration overestimates abundance. In the case of GRB060218/SN2006aj, using the $T_e$ method \cite[see ][]{Wiersema:2007p10082} measure the oxygen abundance of the host as 12+log(O/H$)=7.5$ but when using the PP04 calibration the host sits at 12+log(O/H$)=8.1$. Given that the moderately low PP04 abundance of the host of SN 2007bg somewhat contradicts its extremely low luminosity, it seems that the abundance has been over-estimated. It will be of great value to reobserve the host of SN 2007bg to get a measurement of the oxygen lines required to determine a $T_e$ oxygen abundance. 

\cite{Christensen:2008p10171} have performed the most comprehensive study of the host galaxy of GRB980425/SN1998bw to date, using VIMOS integral field spectroscopy to produce a PP04 abundance map of the galaxy (ESO 184-G82). They find that of all regions in the galaxy the WR region located at a projected distance 0.8 Kpc North-West of the GRB/SN site \citep[originally defined by][]{Hammer:2006p9907} has the lowest oxygen abundance (12+log(O/H$)_{\rm PP04}=8.16${\tiny $\pm0.14$}) and that the GRB/SN region has the second lowest abundance (12+log(O/H$)_{\rm PP04}=8.30${\tiny $\pm0.14$}). We use the abundance measured at the SN/GRB site to plot the host galaxy of GRB980425/SN1998bw in Figure \ref{fig:PP04}. Note that the WR region is the only region in the galaxy where the temperature sensitive [O {\tiny III}] $\lambda$4363 line is detected, allowing \cite{Christensen:2008p10171} to derive a $T_e$ abundance of 12+log(O/H$)=8.53${\tiny $\pm$0.10}. Considering that the abundance at the SN/GRB site is greater than that of the WR region, and taking the uncertainties into account, 12+log(O/H$)=8.43$ can be used as a lower-limit to the $T_e$ abundance at the SN/GRB site. This provides confidence that the PP04 calibration has {\it not} overestimated the abundance in the case of GRB980425/SN1998bw.

In addition, we also plot XRF 080109/SN 2008D in Figure \ref{fig:PP04} using the host galaxy line flux measurements from \cite{Tanaka:2009p10410}. As the source of the X-ray emission of this object is still unclear, with some claiming the SN shock breakout as the source \citep{Soderberg:2008p8522,Chevalier:2008p224} and others a mildly relativistic jet \citep{Li:2008p11070,Mazzali:2008p10760}, we shall not draw conclusions from the position of the object on the plot.

The dotted line indicates the apparent divide suggested by \cite{Modjaz:2008p87} between SNe Ic-BL with and without an associated GRB. The green circles represent 25 nearby, low-luminosity dwarf galaxies used by \cite{Lee:2006p6491} to determine a luminosity-metallicity relationship for this population, itself represented by the green dashed line. The grey data points represent the SDSS-DR4 sample of local star-forming galaxies ($0.005 < z < 0.25$) calibrated to the \cite{Pettini:2004p217} abundance scale. This sample is a super-set of the \cite{Tremonti:2004p173} SDSS-DR2 data and has kindly been made publicly available\footnote{http://www.mpa-garching.mpg.de/SDSS/DR4/}. The line-fluxes used to measure the oxygen abundances of the SDSS data have been corrected for Galactic foreground extinction \citep{Schlegel:1998p99} and for stellar Balmer absorption. Internal extinction is accounted for by iteratively reporting the observed H$\alpha$/H$\beta$ Balmer decrement to an intrinsic value of 2.86 \citep[case {\it B} recombination, ][]{Osterbrock:1989p8676} and applying the \cite{Cardelli:1989p3748} Galactic extinction curve ($R_V=3.1$). The SDSS petrosian magnitudes are $k$-corrected to $z=0$ and converted to the $B$-band using the transformation of \cite{Smith:2002p2474}.

In the latest spectrum of SN 2007bi the narrow emission lines of [O {\tiny II}] $\lambda$3727, [O {\tiny III}] $\lambda\lambda$4959, 5007, and H$\beta$ attributed to the host galaxy are clearly visible above the continuum of the SN. \cite{Pagel:1979p8297} first used this set of lines to estimate an oxygen abundance (R$_{23}$ strong-line ratio). Many other authors have since developed this method, including \cite{McGaugh:1991p7985}. The non-detection of a narrow [N {\tiny II}] $\lambda 6583$ emission line from the host galaxy of SN 2007bi and the knowledge that the host is relatively sub-luminous allows us to place the host on the {\it metal-poor} rather than the {\it metal-rich} branch of the double valued R$_{23}$ ratio. Subtracting off the SN continuum at the location of each of each of the relevant emission lines and measuring the fluxes via fitting Gaussian profiles to each line we estimate 12+log(O/H$)_{\rm M91}=8.15${\tiny $\pm0.13$}, where a conservative 20\% error is included in the measurement of the line fluxes to account for inaccuracies in subtracting off the SN continuum. The placement of SN 2007bi on the \cite{McGaugh:1991p7985} (M91) abundance scale is shown in Figure \ref{fig:M91}. Data points are as described in Figure \ref{fig:PP04}. The description of the M91 calibration found in \cite{Kobulnicky:1999p7983} is used to determine the SDSS galaxy abundances, using the [N {\tiny II}]/[O {\tiny II}] ratio to break the degeneracy in the R$_{23}$ ratio as advised by \cite{Kewley:2008p31}.

Figures \ref{fig:PP04} and \ref{fig:M91} are reconstructions of the PP04 and the M91 oxygen abundances vs absolute magnitude plots produced by \cite{Modjaz:2008p87} (their Figure. 6). Type Ic SNe 2007bg and 2007bi were not associated with GRBs but both fall within the same low-metallicity environment as inhabited by GRB/SNe Ic-BL.

\section{Discussion}\label{Discussion}

\subsection{Low-metallicity SNe Ic-BL without associated GRBs}


\begin{figure}[t]		
\begin{center}
\includegraphics[scale=0.52]{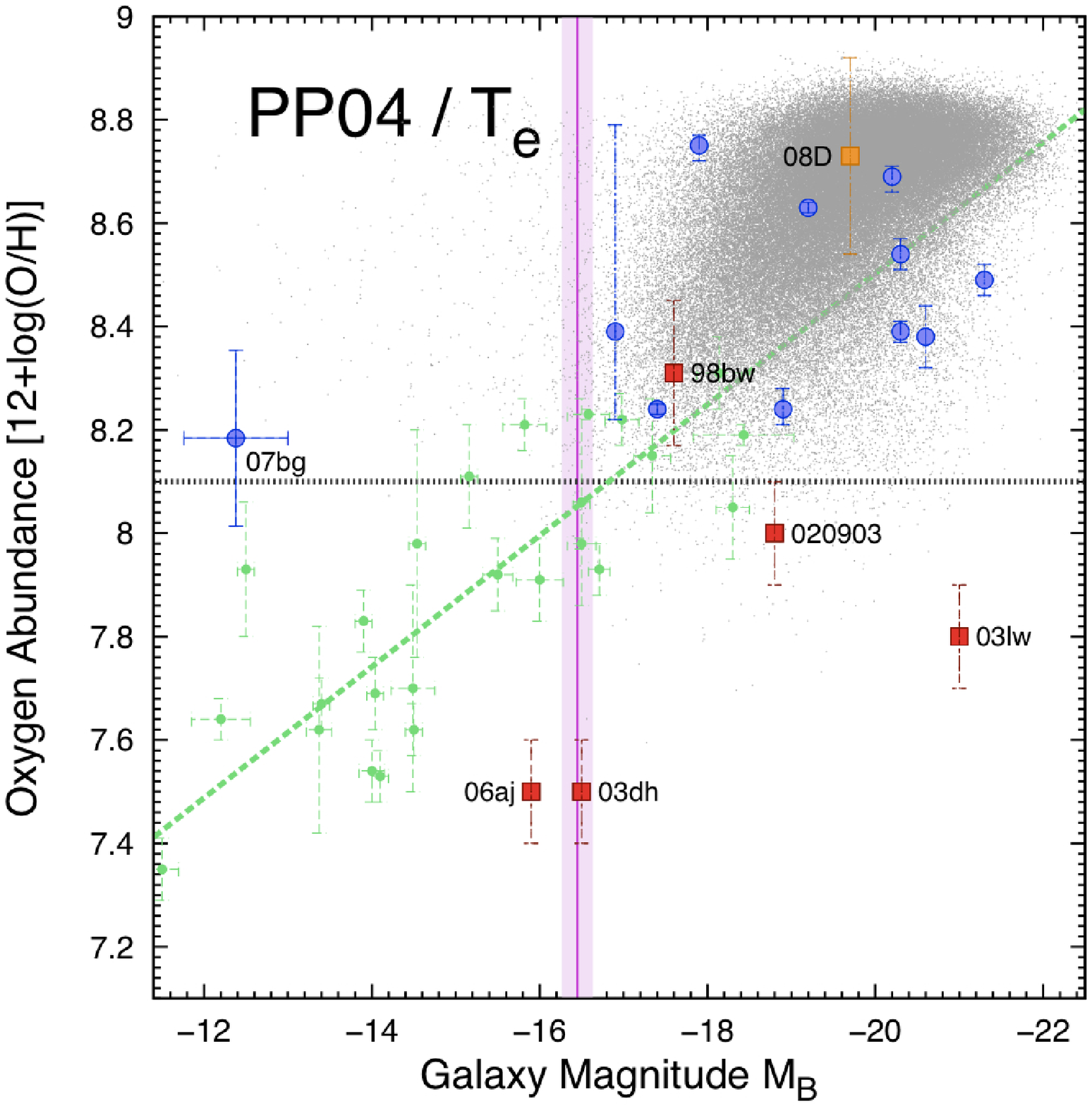}
\caption{Reconstruction of the \cite{Pettini:2004p217} oxygen abundance vs absolute magnitude plot found in Figure 6. of \cite{Modjaz:2008p87}. The {\it red squares} represent those SNe Ic-BL associated with a GRB \citep[now also including GRB980425/SN1998bw with a PP04 abundance measured by][the most recent study of the host galaxy.]{Christensen:2008p10171} and the {\it blue circles} represent those SNe Ic-BL with no associated GRB (now including SNe 2007bg). The {\it dotted line} indicates the apparent divide suggested by \citeauthor{Modjaz:2008p87} between those SNe with and without associated GRBs. The {\it grey points} are SDSS DR4 local star-forming galaxies \citep{Tremonti:2004p173}. The {\it green points} are the 25 dwarf irregular galaxies used by \cite{Lee:2006p6491} to derive the luminosity-metallicity relationship ({\it green dashed line}) for low-mass galaxies. The {\it solid pink} line indicates the magnitude of the host galaxy of SN 2007bi with the shaded area either side representing the uncertainties.}
\label{fig:PP04}
\end{center}
\end{figure}

\begin{figure}[t]		
\begin{center}
\includegraphics[scale=0.52]{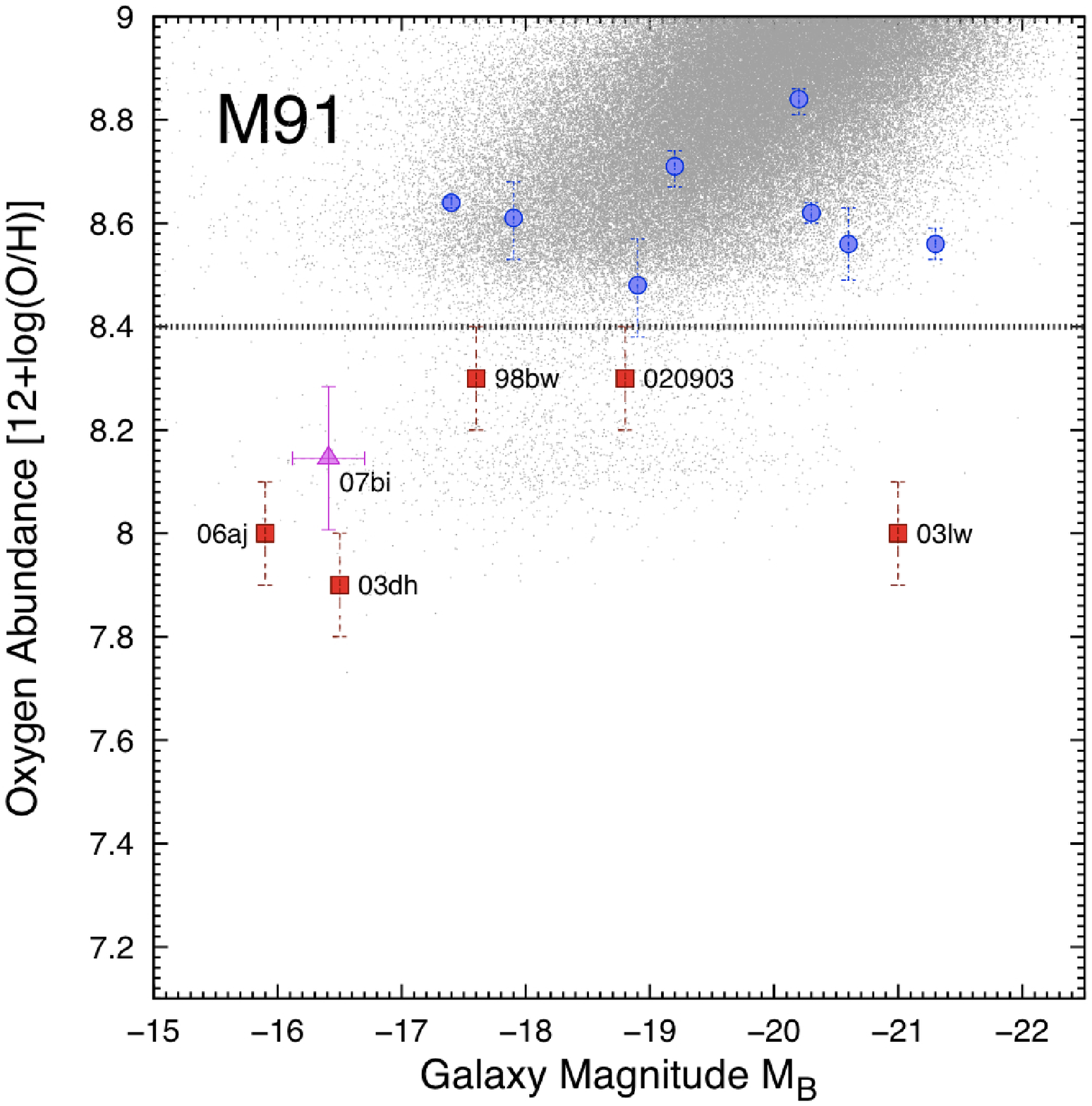}
\caption{Reconstruction of the \cite{McGaugh:1991p7985} oxygen abundance vs absolute magnitude plot found in Figure 6. of \cite{Modjaz:2008p87}. The {\it red squares} represent those SNe Ic-BL associated with a GRB and the {\it blue circles} represent those SNe Ic-BL with no associated GRB. The {\it dotted line} indicates the apparent divide suggested by \citeauthor{Modjaz:2008p87} between those SNe with and without associated GRBs. The {\it grey points} are SDSS DR4 local star-forming galaxies \citep{Tremonti:2004p173}. SN 2007bi is represented by the {\it pink triangle}.}
\label{fig:M91}
\end{center}
\end{figure}

From measurements of the oxygen abundances of the host-galaxy of SNe 2007bg and 2007bi, we find that these objects lie within similar low-metallicity environments as GRB/SNe Ic-BL (see Figures \ref{fig:M91} and \ref{fig:PP04}). The discovery that SN 2007bg, a non-GRB/SN Ic-BL, occupied a similar metal-poor environment as GRB/SNe Ic-BL suggests that there must be secondary distinguishing characteristic that can be used to differentiate between progenitors of GRB/SN Ic-BL and non-GRB/SN Ic-BL besides metallicity. The `Collapsar' is the favoured model for describing the creation of a long-duration GRB \citep{Woosley:1993p8292,Woosley:2006p167} and requires a massive progenitor star which is (a) stripped of both its H and He envelopes and (b) is rotating fast enough so that upon collapse the resulting black hole remnant forms an accretion disk and powers the release of highly relativistic jets along its polar axes. As the progenitor lacks both H and He it is thought to be a Wolf-Rayet star (WR), specifically a WC or WO star \citep{Georgy:2009p9164}, and the resulting associated SN is a Type Ic (confirmed observationally as the only type of SN found in association with a GRB). Low-metallicity must be invoked for this model to ensure that the massive progenitor suffers minimal mass-loss and retains the majority of its initial angular momentum. Using rotating single star models \cite{Georgy:2009p9164} find that if they limit the progenitors of GRBs to low-metallicity WC/WO stars they over-estimate the GRB/CCSNe ratio suggesting, as we do, that metallicity is not the only physical characteristic that distinguishes GRB/SNe Ic from `typical' SNe Ic. They suggest that it must only be the fastest rotating metal-poor progenitors that give rise to a GRB \citep[see ][]{Yoon:2005p9292}.

Given that stellar mass-loss is known to scale with metallicity it is reasonable to assume that in the most metal deprived environments we can expect to see little, if any, evidence of a WR population. In these environments mass-loss rates are so low that even the most massive stars will fail to evolve to a WR phase if main-sequence stellar winds are the only means of stripping their H-envelopes. However, contrary to speculation, evidence of WR populations has been found in these extreme environments, namely the blue compact dwarf galaxies I Zw 18 \citep{Brown:2002p8628,Izotov:1997p9555,Legrand:1997p9563} and SBS 0335-052E \citep{Papaderos:2006p213}. Besides main-sequence stellar winds, there must be another mechanism via which a star can lose its H-envelope enabling WR populations to reside within these extremely metal-poor galaxies. Two viable mechanisms are (i) a quasi-chemically homogeneous evolution of fast rotating massive stars and (ii) stripping via a close interacting binary companion. Chemical mixing due to rotation is thought to be more efficient at low-metallicities (weakness of the main-sequence stellar winds allows angular momentum to be conserved) and some massive stars may undergo a quasi-chemically homogeneous evolution, evolving to become WR stars without the need of extensive mass-loss \citep{Maeder:1987p9570,Yoon:2006p7,Woosley:2006p164,Meynet:2007p9577}. 

With their binary models \cite{Eldridge:2008p9170} have shown that it is possible within a close interacting binary scenario to strip a star of its outer layers to create WR stars, even low-mass WR stars at low-metallicities. Analysing pre-explosion images of the site of Type Ic SN 2002ap, \cite{Crockett:2007p8658} fail to detect a progenitor star. Upon using the luminosity limits imposed by these images they are able to rule out that the progenitor was a single WR star (assuming standard mass-loss rates). They conclude that the progenitor was most likely a 15-20$M_\odot$ star stripped by a binary companion to become a 5$M_\odot$ WR star.
Other mechanisms via which a massive star can lose its H-envelope in a low-metallicity environment may include LBV type outbursts \citep[e.g.][]{Pastorello:2007p236,Foley:2007p9876} and pulsational pair-instability events \citep{Woosley:2007p96}.

\subsection{Confirming the heterogeneity of SNe Ic}

Both SNe 2007bg and 2007bi help to confirm the heterogeneity of SNe Ic. \cite{Richardson:2002p111} and \cite{Richardson:2006p72} have tentatively suggested a case for two groups of stripped-envelope SNe (SNe IIb, Ib and Ic); one {\it normal}-luminosity set with mean M$_V$ = $-$17.7 and an {\it over}-luminous set with mean M $_V$ = $-$20.08. In Figure \ref{RLC-comp-plot} SN 2007bg can be seen to fall intermediate to the {\it over}-luminous SNe 1998bw and 2003jd and the cluster of {\it normal}-luminosity SNe Ic. On the other hand SN 2007bi outstrips the peak magnitude of the over-luminous SN 1998bw by $\sim$2 mags and seems to follow the decay rate of $^{56}$Co to $^{56}$Fe (0.0098 mags d$^{-1}$) for the whole of its observed photometric evolution. This suggests that SNe Ic as a group span a range of peak luminosities and cannot be compartmentalised into two broad groups according only to their luminosities.

\cite{Taubenberger:2006p4726} compare the photometric and spectroscopic evolutions of SNe 1994I and 1997ef to illustrate that the decline rate of the light curves of SNe Ic may be related to the speed of spectral evolution, with a slower decline rate in the light curve related to a slower spectroscopic evolution (related to the photon diffusion time). SN 1990B \citep{Clocchiatti:2001p6126} presented a case clearly contrary to this line of thought with a slow decline in light curve evolution but a fast spectroscopic evolution. However, the occurrence of SNe 2007bg and 2007bi now add weight to this prediction and objects like SN 1990B may simply be atypical to a general rule regarding the evolution of SNe Ic. At early-times SN 2007bg evolved extremely fast photometrically, matching the fast-paced decline of SN 1994I and giving it the one of the fastest post-maximum decline rates of all SNe BL-Ic \citep[for a comparison of decline rates see][]{Sahu:2009p16}. This fast decline in the light curve combined with high expansion velocities also give SN 2007bg a high kinetic energy to ejected mass ratio; $E_K/M_{ej}\sim2.7$. SN 2007bg also seemed to be evolving relatively fast spectroscopically with indications of transitioning to the nebular phase as early as + 58 d (see Section \ref{spec_evol}). SN 2007bi sits at the opposite extreme concerning the relationship between the light curve and spectral evolution. The spectral evolution is extremely slow and the object is still not fully nebular at + 367 d, which complements the remarkably slow decline in its light curve.

\subsection{SN 2007bi: Pair-instability or massive core collapse?}\label{pisn}

Compared with other SNe Ic the slow decline rate of SN 2007bi is most closely matched by SNe 1997ef and 1997dq \citep{Mazzali:2004p5906} both of which indicated little, if any, $\gamma$-ray escape for the duration of their observed light curves. This slow decline rate of the light curve, indicative of close to total $\gamma$-ray trapping, suggests that SN 2007bi had a huge ejected mass. As the light curves of SNe Ic are powered solely by the radioactive decay of $^{56}$Ni, from the luminosity of SN 2007bi we must draw the conclusion that an extraordinary amount of $^{56}$Ni was synthesized in the explosion. From the pseudo-bolometric light curve we estimate $M_{Ni}=3.5-4.5$M$_{\odot}$. From modeling of a fully-nebular spectrum \cite{Gal-Yam:2009} measure $M_{Ni}=4-6$M$_{\odot}$, the largest amount of $^{56}$Ni measured in the ejecta of a SN. Evidence of this copious amount of $^{56}$Ni is notable in the abundance of iron present in the spectra of SN 2007bi (see Figure \ref{fig:Ia_comp_fig}). 

The combination of the remarkable luminosity, the broad peak and large amount of $^{56}$Ni synthesised in the explosion could be explained with the model of a pair-instability supernova \citep{Barkat:1967p9753,Rakavy:1967p9756,Bond:1984p9774}. These events are thought to occur when the temperature in the core of a very massive stars exceeds temperatures of $\sim$$10^9$K allowing electron-positron pair production to occur. Thermal energy is converted into the rest-masses of these newly created particles resulting in a loss of pressure leading to hydrostatic instability \citep{Heger:2002p245,Scannapieco:2005p128,Heger:2003p30}. Explosive oxygen burning ensues which may completely disrupt the entire star. Light curves of these events have been simulated by \cite{Scannapieco:2005p128} and look remarkably similar to that of SN 2007bi, with a slow rise to maximum, high-luminosity, broad peak followed by a slow decline.

\cite{Langer:2007p228} have used rotating single star models to describe two viable progenitors of pair-instability supernovae (PISNe). The first of these is a slowly rotating H-rich massive ($140 - 260M_{\odot}$) yellow supergiant. These stars suffer mass-loss from main-sequence winds but will retain a He-core massive enough to explode as a PISN if $Z \lesssim Z_{\odot}/3$. The second viable progenitor is a fast-rotating very massive star ($M \gtrsim 80M_{\odot}$) that evolves quasi-chemical homogeneously to produce a massive WR star. This second progenitor option is thought to be limited to $Z \lesssim Z_{\odot}/1000$ and therefore only to Population III stars.

It is hard to reconcile SN 2007bi with these models as although the SN satisfies the metallicity constraint of the first progenitor option, it is not H-rich as required by this model and others \citep{Scannapieco:2005p128}. The progenitor of SN 2007bi was certainly a massive WR star, but it fails to satisfy the extremely metal-poor constraint of the second progenitor option.

Faced with the similar problem of having to resolve how SN 1999as \citep{Knop:1999p6878,Hatano:2001p4590} and SN 2006gy \citep{Smith:2007p158,Ofek:2007p15} were able to synthesise such large masses of $^{56}$Ni, \cite{Umeda:2008p6697} evolved metal-poor stellar models from the main-sequence up until Fe core collapse to determine how much $^{56}$Ni core-collapse SNe are capable of producing. $^{56}$Ni is produced via the explosive burning of Si in both core-collapse and thermonuclear SNe. \cite{Umeda:2008p6697} find that the amount of $^{56}$Ni synthesised in core-collapse events is related to both the density structure of the progenitor star and the energy of the explosion. A centrally concentrated star with a large explosion energy has the ability to convert a large amount of Si into $^{56}$Ni. \citeauthor{Umeda:2008p6697} are able to model an ejected $^{56}$Ni mass of $3-6$$M_{\odot}$ for a star of initial mass $50-100M_{\odot}$ and an explosion energy of $25-75\times10^{51}$ ergs. These models predict that a huge amount of oxygen is also ejected in the explosions of these massive stars. For example a star with an initial mass of 90$M_{\odot}$ and an explosion energy of 30$\times10^{51}$ is expected to eject $\sim$18$M_{\odot}$ of oxygen. The results of these core-collapse models are consistent with our findings for SN 2007bi but do not require the almost metal-free environments necessitated by the PISNe models. When considering these core-collapse models, our speculation that the $E_K/M_{ej}$ ratio for SN 2007bi is relatively low (see Section \ref{ejected_mass}) points towards a star of higher intital mass which requires a lower explosion energy to produce the amount of $^{56}$Ni observed in SN 2007bi.

Interaction of the SN ejecta with a dense, optically thick shell of material cannot be ruled out as an alternative (or an additional) mechanism for generating the huge luminosity associated with SN 2007bi. This scenario has been invoked to explain the over-luminous Type IIL SN 2008es \citep{Gezari:2009p2983,Miller:2009p2870}.

It remains to be seen which model, the pair-instability explosion, the massive core-collapse and/or interaction with a dense, optically thick shell best describes the observational characteristics of SN 2007bi.  In each case an initially massive star which has lost its H and He layers is required, which needs to occur in a metal poor regime. Further work on the nucleosynthesis of other elements (O, Ca, Na for example) in each model and
the rate of these rare events (in future surveys) may be the best way to distinguish between the explosion physics.

\section{Summary and Conclusions}\label{Conclusions}

We have presented {\it BVRI} photometric and optical spectroscopic data for SN 2007bg and SN 2007bi. Neither SNe 2007bg nor 2007bi were found in association with a GRB, but from estimates of the metallicities of their host-galaxies they are found to inhabit similar low-metallicity environments to GRB/SNe Ic. SN 2007bg is found to be hosted by an extremely sub-luminous galaxy of magnitude $M_{B}=-12.44${\tiny$\pm0.78$} mag with a measured oxygen abundance of 12+log(O/H$)_{\rm PP04}=8.18${\tiny $\pm0.17$}. SN 2007bg also displays one of the fastest post-maximum decline rates of all SNe BL-Ic. This fast light curve decline combined with high expansion velocities gives SN 2007bg a high kinetic energy to ejected mass ratio ($E_K/M_{ej}\sim2.7$).

   Reaching a peak magnitude of M$_{R}\sim-21.3$ mag, the light curve of SN 2007bi displays a remarkably slow decline, following the decay rate of $^{56}$Co to $^{56}$Fe throughout the course of its observed lifetime. SN 2007bi also displays an extreme longevity in its spectral evolution and is still not fully nebular at approximately one year post-maximum. From modeling the bolometric light curve of SN 2007bi we estimate a total ejected $^{56}$Ni mass of $M_{Ni}=3.5-4.5{\rm M}_\odot$, the largest $^{56}$Ni mass measured in the ejecta of a SN to date. The spectra are characterised by the presence of strong oxygen and iron lines, suggesting the explosion of a massive stellar core which synthesised a remarkable amount of $^{56}$Ni. A strong case can be made that SN 2007bi was a pair-instability supernova, but given that it is hosted by only a moderately low-metallicity galaxy we suggest that a less-exotic core-collapse mechanism may have been the cause of this explosion.
   
   In the future we hope to obtain a spectrum of the host-galaxy of SN 2007bi, now that the SN has faded, in order to more accurately measure the oxygen abundance. It will also be of value to reobserve the host of SN 2007bg to get a measurement of the oxygen lines required to determine a $T_e$ oxygen abundance. It may be that the low-metallicities of the host-galaxies of these two SNe Ic have been one of the main parameters contributing to their peculiarity. With the advent of current and future all-sky surveys such as Pan-STARRS, Skymapper, PTF and RSVP we are certain to find many more of these strange and interesting objects hosted in low-metallicity environments \citep[see][]{Young:2008p8179}.

\begin{acknowledgements}

This work, conducted as part of the award ``Understanding the lives of massive stars from birth to supernovae" (S.J. Smartt) made under the European Heads of Research Councils and European Science Foundation EURYI (European Young Investigator) Awards scheme, was supported by funds from the Participating Organisations of EURYI and the EC Sixth Framework Programme. SJS and DRY thank the Leverhulme Trust and DEL for funding. SM acknowledges funding from the Academy of Finland (project 8120503). We thank Evan Skillman, Henry Lee, Fabio Bresolin and Max Pettini for discussions on metallicity measurements.

This paper is based on observations made with the following  facilities: the European Southern Observatory 8.2-m Very Large Telescope (Chile) and 8.1-m Gemini Telescope North (Hawaii) under program IDs 082.D-0292(A) and GN-2009A-Q-5 respectively, the 2.0-m Liverpool Telescope of the Liverpool John Moores University (La Palma), the 2.0-m Faulkes Telescope North of the Las Cumbres Observatory Global Telescope Network (Hawaii), the 1.82-m Copernico Telescope of the Asiago Observatory (Italy), the 3.58-m Italian National Telescope Galileo operated by the Fundación Galileo Galilei of the INAF (La Palma), the 4.2-m William Herschel Telescope and the 2.5-m Isaac Newton Telescope operated by the Isaac Newton Group (La Palma). We are grateful to the staff at all of the telescopes for their assistance.

\end{acknowledgements}

\bibliography{bibtex/mybib}

\end{document}